\documentclass[journal=jacsat,manuscript=article]{achemso}
\usepackage[version=3]{mhchem} 

\author{D.~Ivaneyko$^{a,b}$}
\email{ivaneiko@ipfdd.de}
\author{V.~Toshchevikov$^{b,c}$}
\author{M.~Saphiannikova$^{b}$}
\author{G.~Heinrich$^{b,a}$}
\affiliation{
$^a$Institute of Materials Science, Technical University of Dresden, Helmholtz Str. 7, 01069 Dresden, Germany,\\
$^b$Leibniz Institute of Polymer Research Dresden, Hohe Str. 6, 01069 Dresden, Germany,\\
$^c$Institute of Macromolecular Compounds, Russian Academy of Science, Bolshoi Prospect 31, V.O., Saint-Petersburg, 199004, Russia}

\title[MSEs]
{Magneto-sensitive elastomers in a homogeneous magnetic field: a regular rectangular lattice model}

\begin{document}
\begin{abstract}
A theory of mechanical behaviour of the magneto-sensitive elastomers is developed in the framework of a linear elasticity approach. Using a regular rectangular lattice model, different spatial distributions of magnetic particles within a polymer matrix are considered: isotropic, chain-like and plane-like. It is shown that interaction between the magnetic particles results in the contraction of an elastomer along the homogeneous magnetic field. With increasing magnetic field the shear modulus for the shear deformation perpendicular to the magnetic field increases for all spatial distributions of magnetic particles. At the same time, with increasing magnetic field the Young's  modulus for tensile deformation along the magnetic field decreases for both chain-like and isotropic distributions of magnetic particles and increases for the plane-like distribution of magnetic particles.
\end{abstract}

\section{Introduction}

Magneto-sensitive elastomers (MSEs) are a class of smart materials, whose mechanical behaviour can be controlled by application of an external magnetic field. Recently, MSEs as well as magnetic gels (ferrogels) and magnetorheological fluids have been utilized in applications with fast switching processes. In particular, MSEs have been used in controllable membranes, rapid response interfaces designed to optimize mechanical systems and automobile applications such as stiffness tunable mounts and suspension devices.$^[$\cite{Carlson00,Lin08}$^]$ MSEs typically consist of micron-sized iron particles dispersed within an elastomeric matrix, which is highly cross-linked, having the values of Young's modulus about $E\sim10^6 Pa$. The particles are separated by the polymer matrix and are fixed in their positions. In this respect, MSEs differ from ferrogels, in which the polymer matrix is usually only weakly cross-linked, having the values of Young's modulus about $E\sim10^4 Pa$.$^[$\cite{Wood11}$^]$ The typical size of the magnetic particles in ferrogels is of the order of 10 $nm$, that is much smaller than the mesh size of the polymer network.$^[$\cite{Rosenweig85,Zrinyi98,Jarkova03,Wood11}$^]$ Therefore, in contrast to the MSEs, the magnetic particles in ferrogels can diffuse through the network and build some agglomerates.

The spatial distribution of magnetic particles in a magneto-sensitive elastomer can be either isotropic (so-called elastomer-ferromagnet composites) or anisotropic (so-called magnetorheological elastomers),$^[$\cite{Zhou04}$^]$ depending on whether they have been aligned by an applied magnetic field before the cross-linking of the polymer. If the constant magnetic field is applied to a polymer melt with magnetic particles, one obtains after cross-linking the chain-like structures formed by the particles.$^[$\cite{Filipcsei07}$^]$ Recently, the MSEs with plane-like spatial distributions of particles have been synthesised using the magnetic fields with rotating vector of the magnetic strength or a strong shear flow before the cross-linking procedure.$^[$\cite{Kulichikhin09}$^]$

The main object of investigations in many experimental,$^[$\cite{Bednarek99,Coquelle05,Martin06,Guan08}$^]$ theoretical$^[$\cite{Raikher00,Borcea01,Kankanala04,Diguet10}$^]$ and simula\-tion$^[$\cite{Coquelle06,Stepanov08,Raikher08}$^]$ studies was the effect of the shape change of MSEs under magnetic field (magnetostriction effect). At the same time, the effect of the magnetic field on the mechanical moduli of MSEs have been studied not so thorough and is one of the main topics under investigation nowadays. The most of the experimental tests of MSEs for tensile and shear deformations indicate the increase of the elastic modulus$^[$\cite{Bellan02,Varga06,Abramchuk07}$^]$ and shear modulus$^[$\cite{Shiga95,Demchuk02,Lokander03,Deng07,Jiang08,Wu09,Bose09,Boczkowska09,Chertovich10}$^]$ with increasing magnetic strength for both isotropic and anisotropic MSEs.

Until now there have been two kinds of theoretical study of MSEs' modulus in a homogeneous magnetic field. On the one side, the mechanical behaviour of MSEs has been analysed by using a continuum-mechanics approach, in which the electromagnetic equations are coupled with the appropriate mechanical deformation equations. Mathematical modelling of mechanical behaviour of MSEs, using different formulations of balance laws and Maxwell's equation for magnetic field have been done for the case of shearing$^[$\cite{Brigadnov03,Dorfmann03,Marvalova08,Tuan09,Ishikawa09}$^]$ or nonlinear deformation.$^[$\cite{Dorfmann03a,Dorfmann04,Dorfmann04a,Dorfmann05}$^]$ In these studies, however, the discrete material properties in MSEs (especially, chain-like and plane-like distributions of particles) have not been taken into account, as the continuum approach assumes homogeneity of the media. Alternatively, simplified microscopic lattice models have been proposed. In particular, one-chain model called quasi-static one dimensional model$^[$\cite{Jolly96,Jolly96a,Davis99}$^]$ and multi-chain model$^[$\cite{Zhu06}$^]$ of MSEs have been developed. In these papers the influence of magnetic field on the shear modulus has been considered only for the chain-like structures without taking into consideration the effect of magnetostriction on the modulus of MSEs.

In the present paper we develop a theory of mechanical behaviour of magneto-sensitive elastomers in a homogeneous magnetic field, taking their microscopic structure explicitly into account. A regular rectangular lattice model of the MSE is proposed which allows us to consider different spatial distributions of magnetic particles inside an elastomer: isotropic, chain-like and plane-like distributions. The condition of  affine deformation that enables the study of equilibrium elongation is considered. The free energy and static mechanical behaviour of the MSE are examined in a homogeneous magnetic field. The dependence of the equilibrium elongation on the strength of magnetic field is considered for different volume fractions of particles and different values of matrix elasticity. Two types of small deformation applied to the MSE exposed to the magnetic field are studied: shear deformation and tensile deformation. The shear and tensile moduli are calculated as functions of the magnetic field taking into account the magnetostriction effect.

\section{Microscopic model of a magneto-sensitive elastomer}\label{S2}

\begin{figure}[!ht]
\centerline{\includegraphics[width=\linewidth]{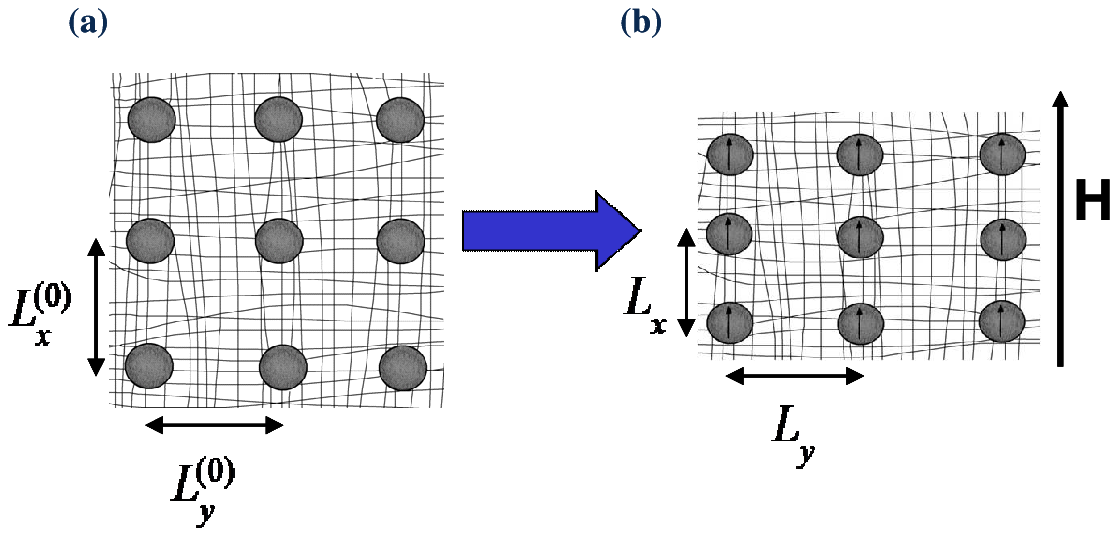}}
\caption{\label{F1} A model of an MSE with magnetic particles arranged on the sites of a regular rectangular lattice, when (a) the external magnetic field $\mathbf{H}$ is turned off, or (b) the external magnetic field $\mathbf{H}$ is turned on.}
\end{figure}

\begin{figure}[!ht]
\centerline{\includegraphics[width=\linewidth]{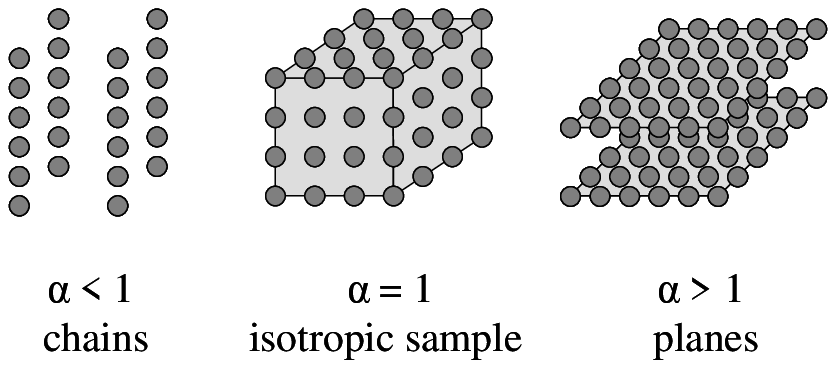}}
\caption{\label{F2} Three different spatial distributions of magnetic particles inside an MSE.}
\end{figure}

To describe the spatial distribution of magnetic particles inside a magneto-sensitive elastomer a lattice model is used, see \ref{F1}. In this model, it is assumed that the magnetic particles are located at the sites of a regular rectangular lattice. In the absence of an external magnetic field, the distances  between neighbouring particles along the $x$-, $y$- and $z$-axes are $L^{(0)}_x$, $L^{(0)}_y$ and $L^{(0)}_z$, respectively. We assume that the distance $L^{(0)}_x$ can differ from the distances $L^{(0)}_y$ and $L^{(0)}_z$: $L^{(0)}_x\neq L^{(0)}_y=L^{(0)}_z$. Furthermore, we introduce a dimensionless parameter $\alpha=L^{(0)}_x / L^{(0)}_y$ in order to describe different spatial distributions of magnetic particles in a polymer matrix: isotropic distribution ($\alpha=1$), chain-like distribution ($\alpha <1$) and plane-like distribution ($\alpha >1$), see \ref{F2}. Under such assumption, the $x$-axis  is the axis of symmetry of an MSE: it lies along the chains in the chain-like structures and is perpendicular to the planes formed by the magnetic particles in the plane-like structures.

\begin{figure}[!ht]
\centerline{\includegraphics[width=10cm]{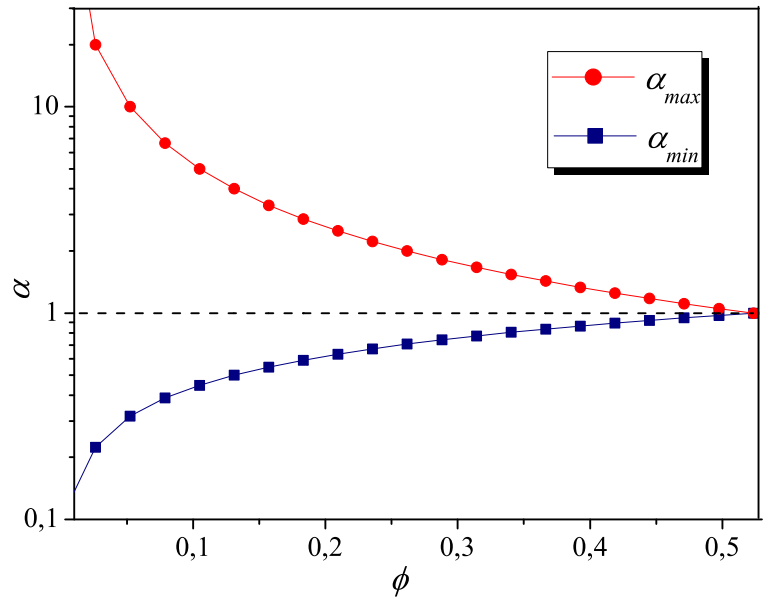}}
\caption{\label{F3} Values of
$\alpha_{min}$ and $\alpha_{max}$ as functions of the volume fraction $\phi$ of magnetic particles.}
\end{figure}

For simplicity we assume that all particles are the same and have a spherical form; $r$ is the radius of particles. The value of $r$ characterizes the average size of particles in a real elastomer. Then, the volume fraction, $\phi$, of the particles is given by:
\begin{equation}\label{A1}
\phi = \frac{\frac{4}{3}\pi r^3}{L^{(0)}_x L^{(0)}_y L^{(0)}_z}.
\end{equation}
Depending on the volume fraction of magnetic particles $\phi$, the parameter $\alpha $ can vary between its minimal and maximal values: $\alpha_{min} < \alpha < \alpha_{max}$. Here we take into account that the particles are rigid and can not penetrate in one another. We obtain the value of $\alpha_{min}$ after substitution of the relation $L^{(0)}_x =2r$ and $L^{(0)}_y=L^{(0)}_x/\alpha=2r/\alpha$ into Equation (\ref{A1}) in the following form:  $\alpha_{min} = (6\phi / \pi)^{1/2}$. Substituting the conditions $L^{(0)}_y =2r$ and $L^{(0)}_x=\alpha L^{(0)}_y=\alpha 2r$ into Equation (\ref{A1}) one can obtain the value of $\alpha_{max}$ as follows: $\alpha_{max} = \pi / 6\phi$. Dependences of $\alpha_{min}$ and $\alpha_{max}$ on volume fraction $\phi$ are presented in \ref{F3}.

Application of a magnetic field induces an average magnetic moment in each particle along the direction of the field. In our work we consider such a configuration when the magnetic field is directed along the axis of symmetry ($x$-axis in \ref{F1}b). The values of the induced magnetic moments in the magnetic particles depend on the material of the particles. Usually, magnetic particles are prepared from pure iron, iron oxide $Fe_2O_3$ or iron-based alloys such as iron-cobalt and mainly carbonyl iron, with typical size of particles being of the order from hundreds nanometres to a few microns. Carbonyl iron particles are nearly pure $Fe$ and have a shape very close to a sphere.$^[$\cite{Promislow96,Arias06}$^]$ A value of carbonyl iron particle density is close to the value of the bulk density of iron, which is about 7.86 $g/cm^3$.$^[$\cite{Park09,Gama10}$^]$

The magnetic particles of micron-sizes have a multi-domain magnetic
structure. Nevertheless, very narrow hysteresis cycles of carbonyl
iron particles were observed which indicates a soft magnetic
behaviour. The dependence $M(H)$ can be described in a good
approximation by the Fr\"{o}hlich--Kennely
equation$^[$\cite{Bossis99,Arias06}$^]$
\begin{equation}
M=\frac{M_{\rm s}(\mu_{ini}-1)H}{M_{\rm s} + (\mu_{ini}-1)H}, \label{MM}
\end{equation}
where $M_{\rm s}$ is saturation magnetization and $\mu_{ini}$ is magnetic permeability of the particles. The magnetization of the particles, $M(H)$, increases with increasing magnetic field and tends to the saturation magnetization, $M_{\rm s}$, when $H\rightarrow \infty$. The saturation magnetization was estimated to be $M_{\rm s}\approx1582~kA/m$ and magnetic permeability $\mu_{ini} \approx 21.5$ for carbonyl iron particles with the average diameter of $470\pm180~nm$.$^[$\cite{Arias06}$^]$ Similar values were obtained for particles of the size of $2~\mu m$: $M_{\rm s} = 1990~kA/m$ and $\mu_{ini} = 132$.$^[$\cite{Bossis99}$^]$ Also, no loops of hysteresis were observed even for the iron powders with particles of the same size ($2~\mu m$) as well as for MSEs synthesized on the base of this powder.$^[$\cite{Abramchuk07}$^]$ In our further  considerations instead of the strength of magnetic field, $H$, we will use the magnetization, $M$, which is the one-to-one function of $H$ for existing MSEs. Thus, our theory can be formally  applied for superparamagnetic particles as well as for ferromagnetic particles which exhibit very narrow hysteresis cycles.

Interaction between the induced magnetic moments of the particles
leads to pair-wise attraction and repulsion of the magnetic
particles depending on their mutual positions. This interaction
results in a shape change of an MSE. The mechanical response of an
elastomer to the magnetic field is characterized by the value of
the strain $\varepsilon = \Delta l/l$, where $\Delta l$ and $l$
are the elongation and original size, respectively, of an
elastomer along the direction of the magnetic field ($x$-axis).
The condition of constant volume for
elastomers,$^[$\cite{Treloar58,Doi86}$^]$  $L^{(0)}_x L^{(0)}_y
L^{(0)}_z = L_x L_y L_z$, allows us to relate the elongation
ratios $\lambda_x, \lambda_y, \lambda_z$ for the deformation of an
elastomer in the three principal directions as follows:
\begin{eqnarray}
\lambda_x=1+\varepsilon, \hspace{2cm}
\lambda_y=\lambda_z=1/\sqrt{1+\varepsilon}.\label{RA1}
\end{eqnarray}
Here $\lambda_x=L_x/L^{(0)}_x, \lambda_y=L_y/L^{(0)}_y$ and
$\lambda_z=L_z/L^{(0)}_z$. In order to relate displacements of
particles with the macroscopic deformation we use the condition of
affinity of deformation,$^[$\cite{Treloar58,Doi86}$^]$ which can
be written as:
\begin{eqnarray}
(R_{ij})_x&=&(R_{ij}^0)_x\lambda_x=(R_{ij}^0)_x(1+\varepsilon),\label{RA}\\
(R_{ij})_y&=&(R_{ij}^0)_y\lambda_y=(R_{ij}^0)_y(1+\varepsilon)^{-\frac{1}{2}},\label{RB}\\
(R_{ij})_z&=&(R_{ij}^0)_z\lambda_z=(R_{ij}^0)_z(1+\varepsilon)^{-\frac{1}{2}}, \label{RC}
\end{eqnarray}
where $(R_{ij})_{\xi}$ and $(R_{ij}^0)_{\xi}$ are the components of vectors, that separate two magnetic particles after and before deformation, respectively, $(\xi=x,y,z)$.

\section{Free energy}\label{S3}

Mechanical behaviour of an MSE in a magnetic field can be studied
using the equation for the free energy as a function of strain
$\varepsilon$. The free energy consists of two parts: elastic
energy due to entropic elasticity of polymer chains and the
potential energy of magnetic particles. In the linear
approximation for the elasticity of the polymer matrix, the free
energy per unit volume can be written as:
\begin{equation}
F(\varepsilon)=\frac{E_0\varepsilon^2}{2}+u(\varepsilon),\label{FE2}
\end{equation}
where $E_0$ is the Young's  modulus of a filled elastomer. The
value of $E_0$ includes contributions of different possible
effects into the elastic energy appearing under elongation of a
sample: reinforcement of an elastic matrix by the hard particles,
possible adhesion of a polymer matrix on surfaces of hard
particles (glassy-like layers), deformation of interphase domains
of the composite, etc. However, we do not discuss here, how the
value of $E_0$ depends on these effects, since this task is a
special problem in the theory of elasticity for isotropic
reinforced rubbers.$^[$\cite{Vilgis09}$^]$ We use $E_0$ as a
phenomenological parameter of the theory assuming that it can be
extracted from experimental data for elasticity of an MSE in the
absence of the magnetic field. Our task is to describe the
mechanical behaviour of an MSE under application of the magnetic
field and to investigate how this behaviour depends on the value
of $E_0$.

Magnetic part of the free energy, $u(\varepsilon)$, represents the potential energy of the interaction between magnetic particles per unit volume and can be written as: \begin{equation}
u(\varepsilon)=\frac{1}{V}\sum_j
U_j(\{\vec{\textbf{R}}_{ij}(\varepsilon)\}),\label{FE}
\end{equation}
where $\vec{\textbf{R}}_{ij}(\varepsilon)$ is given by Equation (\ref{RA})--(\ref{RC}). In Equation (\ref{FE}) $V$ is the volume of an elastomer and $U_j$ is the potential energy of the $j$-th magnetic particle in the field of all other particles:$^[$\cite{Landau80,Jackson98}$^]$
\begin{equation}
U_j(\{\vec{\textbf{R}}_{ij}(\varepsilon)\}) = - \frac{\mu_r\mu_0}{ 4\pi}\sum_i
\left[\frac{3(\vec{\textbf{m}}_i\cdot\vec{\textbf{R}}_{ij})(\vec{\textbf{m}}_j\cdot\vec{\textbf{R}}_{ij})}{|\vec{\textbf{R}}_{ij}|^5}
-\frac{(\vec{\textbf{m}}_i\cdot\vec{\textbf{m}}_j)}{|\vec{\textbf{R}}_{ij}|^3}\right],\label{UJ}
\end{equation}
where $\mu_0$ is the permeability of the vacuum and $\mu_r$ is the relative permeability of the medium. In the present work we consider an elastomeric matrix to be non-magnetic, therefore everywhere below we set $\mu_r=1$. Here $\vec{\textbf{m}}_i$ and $\vec{\textbf{m}}_j$ are dipole moments of  $i$-th and $j$-th  magnetic particles, $\vec{\textbf{R}}_{ij}$ is the radius vector that joins the $i$-th and $j$-th  magnetic particles.

The value of $U_{j}$ does not depend on the number $j$ due to the translational symmetry for infinite lattice. Thus, we can rewrite Equation (\ref{FE}) as:
\begin{equation}
u(\varepsilon)=c\cdot
U_j(\{\vec{\textbf{R}}_{ij}(\varepsilon)\}),\label{FE1}
\end{equation}
where $c$ is the number of magnetic particles in the unit volume. For calculation of $U_j(\{\vec{\textbf{R}}_{ij}(\varepsilon)\})$ we use that $\vec{\textbf{m}}_i$ and $\vec{\textbf{m}}_j$ are directed along the external field $\textbf{H}$ (the $x$-axis) and their absolute values  are $m_i=m_j= \upsilon_0M$, where $\upsilon_0=\frac{4}{3}\pi r^3$ is the volume of a particle and $M$ is its magnetization. Then Equation (\ref{UJ}) can be rewritten in the form:
\begin{equation}
U_j(\{\vec{\textbf{R}}_{ij}(\varepsilon)\}) = - u_0 \upsilon_0^2\left(\frac{M}{M_{\rm s}}\right)^2\sum_i
\left[\frac{3(\vec{\textbf{R}}_{ij})_x^2-|\vec{\textbf{R}}_{ij}|^2}{|\vec{\textbf{R}}_{ij}|^5}\right],\label{UJ1}
\end{equation}
where we introduce the parameter $u_0$:
\begin{equation}
u_0=\frac{\mu_0 M_{\rm s}^2}{ 4\pi},
\end{equation}
that defines the characteristic energy of magnetic interaction. For $M_{\rm s} \approx 2\times10^6~A/m$  we have $u_0= 4\times10^{5}~Pa$. Below, we will show that mechanical behaviour of an MSE in the magnetic field are determined by the dimensionless parameter $E_0/u_0$, i.e. by the ratio between characteristic values of the elastic and magnetic energies.

\begin{figure}[!ht]
(a) chains\\
\includegraphics[width=7cm]{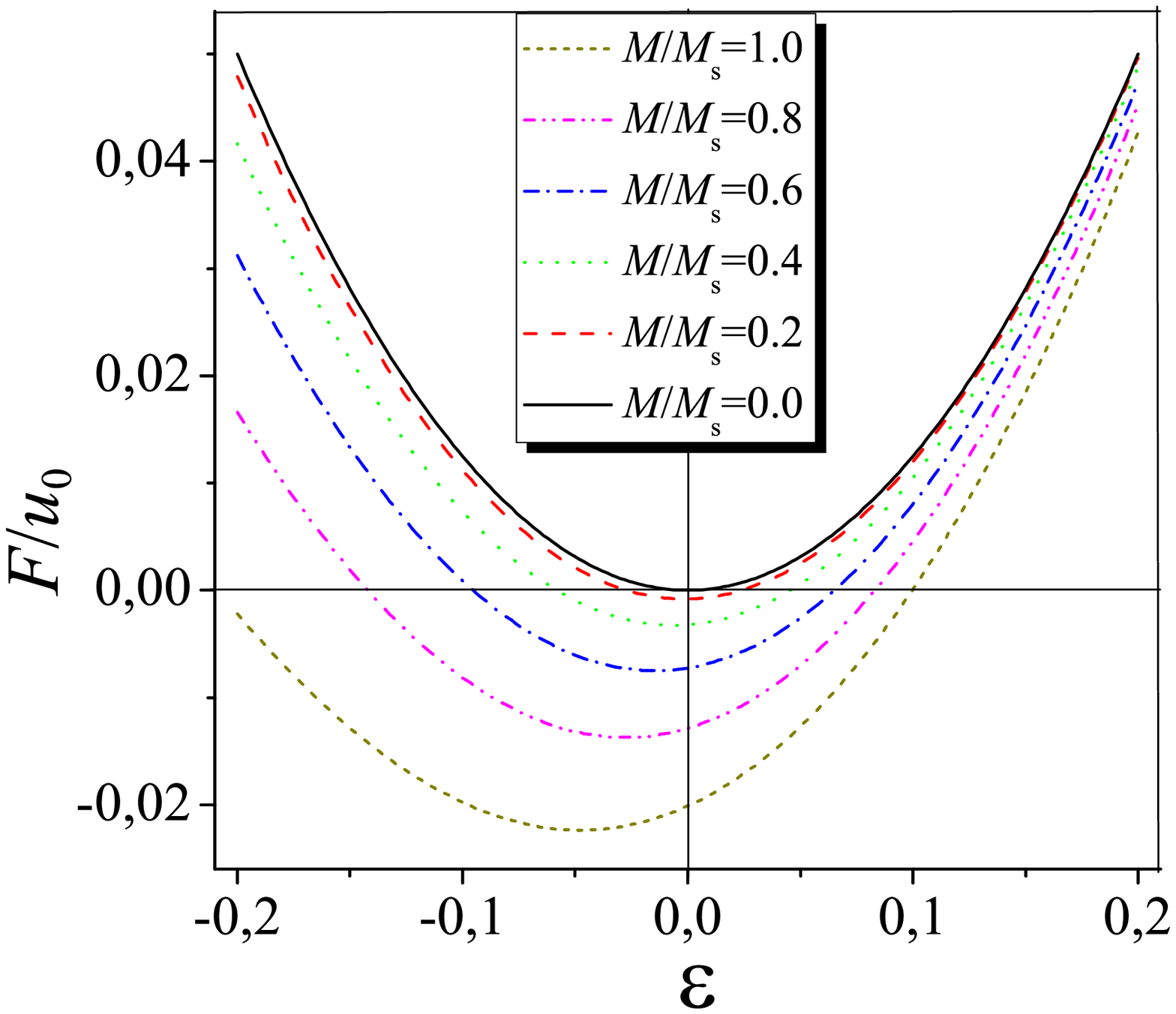}\\
(b) isotropic sample\\
\includegraphics[width=7cm]{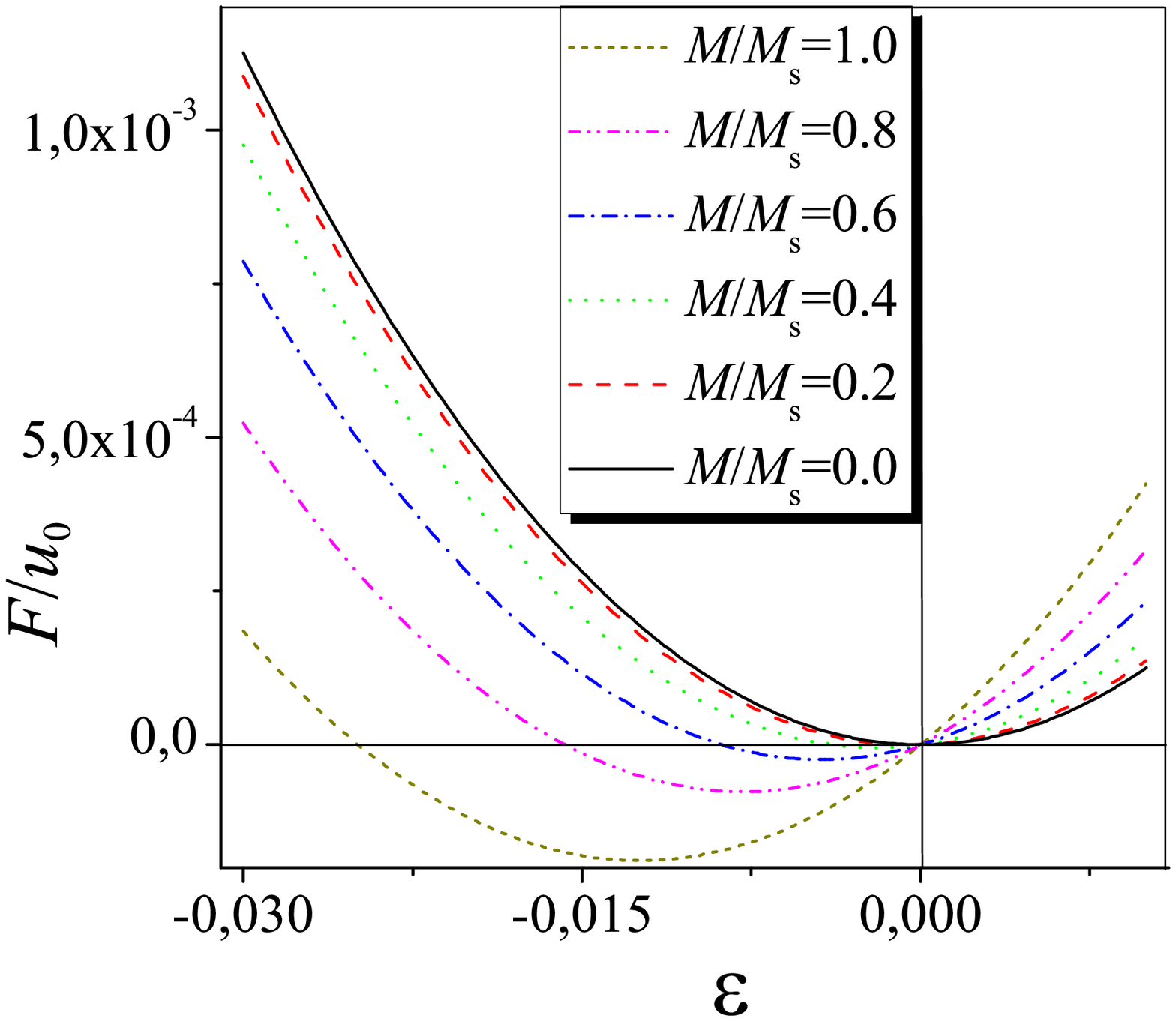}\\
(c) planes\\
\includegraphics[width=7cm]{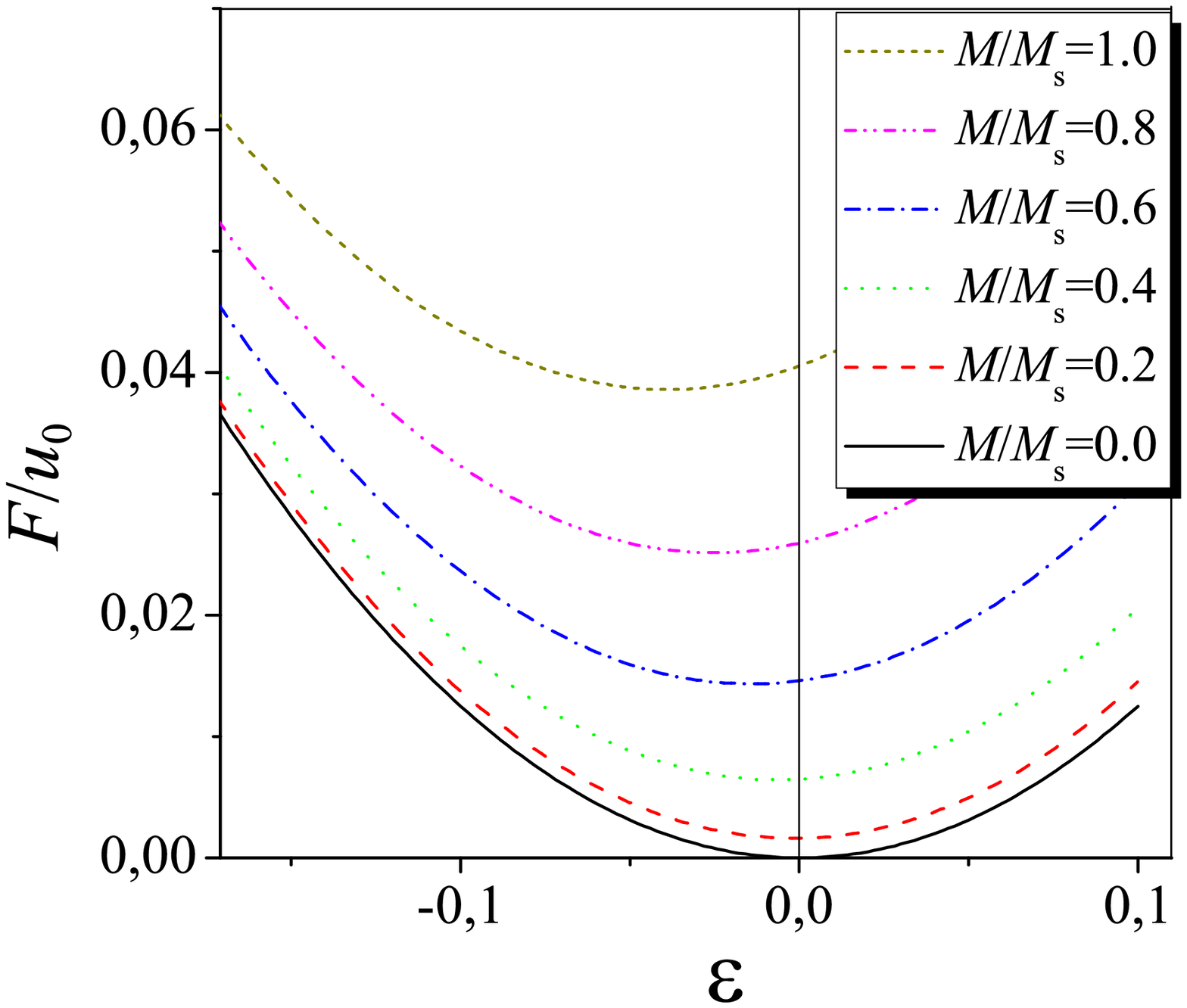}
\caption{\label{F4} Reduced free energy, $F/u_0$, of an MSE as a function of the strain $\varepsilon$ calculated at different values of the reduced magnetization $M/M_{\rm s}$ and at fixed values $\phi=0.05$ and $E_0/u_0=2.5$. The values of the parameter $\alpha$ are given by: (a) Equation (\ref{ALP1}) for the chain-like distributions, (b) $\alpha=1$ for the isotropic distributions, (c) Equation (\ref{ALP2}) for the plane-like distributions.}
\end{figure}

In Equation (\ref{UJ1}) index $i$ numerates the sites of an infinite three-dimensional lattice. The index $i$ can be expressed as a vector $i=(i_x,i_y,i_z)$, where $i_x, i_y, i_z$ are the numbers of cells between  $i$-th and $j$-th particle along the $x$-, $y$- and $z$-axes, respectively. Then the radius vector $\vec{\textbf{R}}_{ij}$ can be presented in the form:
\begin{equation}
\vec{\textbf{R}}_{ij}=(L_x i_x, L_y i_y, L_z i_z).\label{RIJ}
\end{equation}
Using Equation (\ref{RIJ}) and taking into account the relation $L_x/L_y=(\lambda_x/\lambda_y)(L_x^{(0)}/L_y^{(0)})=(1+\varepsilon)^{3/2}\alpha$, Equation (\ref{UJ1}) can be written in the form:
\begin{equation}
U_j(\{\vec{\textbf{R}}_{ij}(\varepsilon)\}) = - u_0
\frac{\upsilon_0^2}{L_y^3}\left(\frac{M}{M_{\rm s}}\right)^2\sum_{\{i_x i_y i_z\}\neq0}
\frac{2\alpha^2(1+\varepsilon)^3i_x^2-i_y^2-i_z^2}{\left[\alpha^2(1+\varepsilon)^3i_x^2+i_y^2+i_z^2\right]^{\frac{5}{2}}}.\label{UJ2}
\end{equation}
Here the sum runs over all sites of rectangular lattice, excluding the point $i_x=i_y=i_z=0$. Substituting Equation (\ref{UJ2}) into (\ref{FE1}) and taking into account the relation $c=1/(L_x L_y L_z)$,  we obtain for $u(\varepsilon)$:
\begin{equation}
u(\varepsilon)= u_0
\phi^2\left(\frac{M}{M_{\rm s}}\right)^2 f(\varepsilon),\label{UJ3}
\end{equation}
where $\phi=\upsilon_0/(L_x L_y L_z)$ is the volume fraction of particles, and the dimensionless function $f(\varepsilon)$ has the following form:
\begin{equation}
f(\varepsilon)=-\alpha(1+\varepsilon)^{\frac{3}{2}}\sum_{\{i_x i_y i_z\}\neq0}
\frac{2\alpha^2(1+\varepsilon)^3i_x^2-i_y^2-i_z^2}{\left[\alpha^2(1+\varepsilon)^3i_x^2+i_y^2+i_z^2\right]^{\frac{5}{2}}}.\label{FFF}
\end{equation}
It can be shown that the sum in the right-hand side converges at any values $\alpha>0$ and $\varepsilon>-1$. In numerical calculations we have approximated the infinite sum in the right-hand side of Equation (\ref{FFF}) by the finite sum: we stop the summation on a finite lattice, for which the increase of the number of layers by unity changes the value of the function $f(\varepsilon)$ no more than by 0.1\%. This procedure provides the value $f(\varepsilon)$ with the errors of about 0.1\%, since the sum in the right-hand side of Equation (\ref{FFF}) converges.  Using Equation (\ref{FE2}), (\ref{UJ3}) and  (\ref{FFF})  we have calculated numerically the free energy as a function of strain $\varepsilon$ at different values of the reduced magnetization $M/M_{\rm s}$. The results are presented in \ref{F4} at fixed values $\phi=0.05$ and $E_0/u_0=2.5$. The value of $E_0/u_0=2.5$ corresponds to the elastic modulus $E_0=10^6~Pa$, when $u_0=4\times10^{5}~Pa$. For the chain-like and plane-like structures of magnetic particles we have chosen the values of the parameter $\alpha$ in such a way, that the initial gap between nearest particles equals the radius of a particle $r$. Then the distance between particles in the chain-like structures is  $L_x^{(0)}=3r$. This gives for the chain-like structures:
\begin{equation}
\alpha_{{\rm ch}}=\sqrt{\frac{81\phi}{4\pi}}.\label{ALP1}
\end{equation}
For the plane-like structures the distance between nearest particles has been chosen as $L_y^{(0)}=3r$, which corresponds to the value of $\alpha$:
\begin{equation}
\alpha_{{\rm pl}}=\frac{4\pi}{81\phi}.\label{ALP2}
\end{equation}
Note, that it was shown experimentally that the chain-like structures contain gaps between particles, these gaps being of the order of the size of particles.$^[$\cite{Coquelle05,Coquelle06}$^]$

One can see from \ref{F4} that application of the magnetic field leads to the shift of the minimum of the free energy to negative values of the strain, $\varepsilon_{\rm eq} <0$, for all considered spatial distributions of particles: isotropic distribution ($\alpha=1$), chain-like distribution ($\alpha <1$) and plane-like distribution ($\alpha >1$). This means that a sample should demonstrate a uniaxial compression along the direction of the magnetic field. Note, that the similar behaviour of $F$ takes place for different volume fractions $\phi$ and for different values of the parameter $E_0/u_0$.

The values of the equilibrium elongation $\varepsilon_{\rm eq}$, that correspond to the minimum of the free energy, as well as the mechanical moduli of the MSE are functions of the magnetic field and of the parameters $\phi$, $E_0$, $u_0$. These dependences are considered in the next sections.

\section{Static mechanical behaviour of MSEs in a homogeneous magnetic field} \label{S4}

\subsection{Equilibrium elongation}

The stress induced by the application of a magnetic field can be
calculated by taking the first derivative of the free energy with
respect to the strain $\varepsilon$:
\begin{eqnarray}
\sigma =-\left. \frac{\partial F}{\partial \varepsilon}\right|_M.
\end{eqnarray}
The equilibrium elongation $\varepsilon_{\rm eq}$ can be found
from the condition $\sigma=0$, that gives the following equation:
\begin{eqnarray}
E_0\varepsilon_{\rm eq}\!+\!u_0
\phi^2\!\left(\!\frac{M}{M_{\rm s}}\!\right)^2\!\alpha\!\sqrt{1+\varepsilon_{\rm eq}}\!\!\!\!\sum\limits_{\{i_x
i_y i_z\}\neq 0}\!\!\!\!\frac{12\alpha^4\!(\!1\!+\!\varepsilon_{\rm eq}\!)^6i_x^4\!-\!30\alpha^2i_x^2\!(\!1\!+\!\varepsilon_{\rm eq}\!)^3\!(\!i_y^2\!+\! i_z^2\!)\!+\!3(\! i_y^2\!+\!i_z^2\!)^2}{2\left[\alpha^2(1+\varepsilon_{\rm eq})^3i_x^2+i_y^2+i_z^2\right]^{\frac{7}{2}}}\!=\!0.\label{EE}
\end{eqnarray}
Dividing both the left- and right-hand sides of Equation
(\ref{EE}) by the factor $u_0$, one can see that the equilibrium
elongation $\varepsilon_{\rm eq}$ depends on the elastic modulus
$E_0$ and on the magnetic parameter $u_0$ through their
dimensionless ratio $E_0/u_0$. We note that $\varepsilon_{\rm eq}$
is an even function of $M$, since the transformation $M
\rightarrow -M$ does not change the solution of Equation
(\ref{EE}) with respect to the parameter $\varepsilon_{\rm eq}$.

\begin{figure}[!ht]
(a) chains\\
\includegraphics[width=7cm]{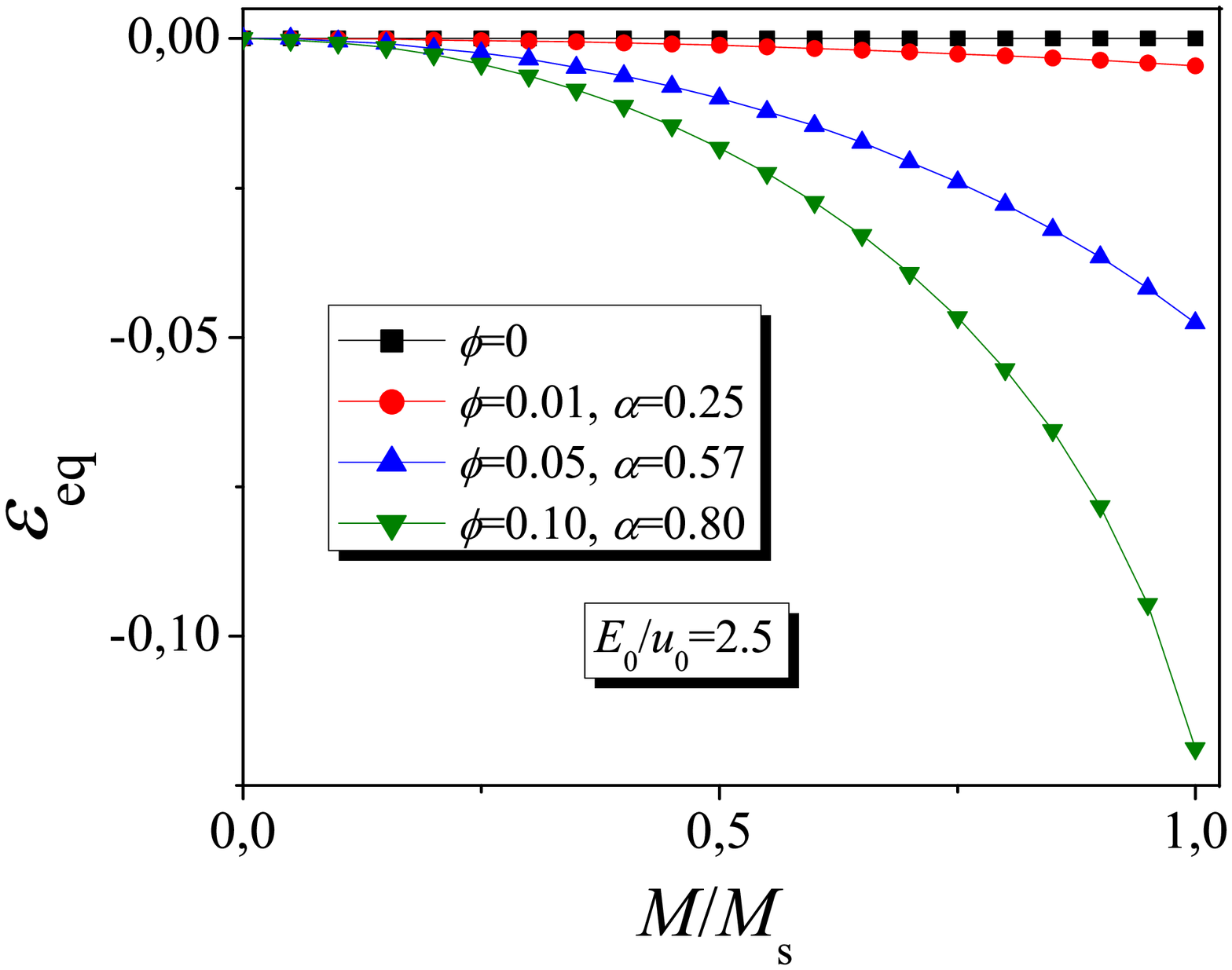}\\
(b) isotropic sample\\
\includegraphics[width=7cm]{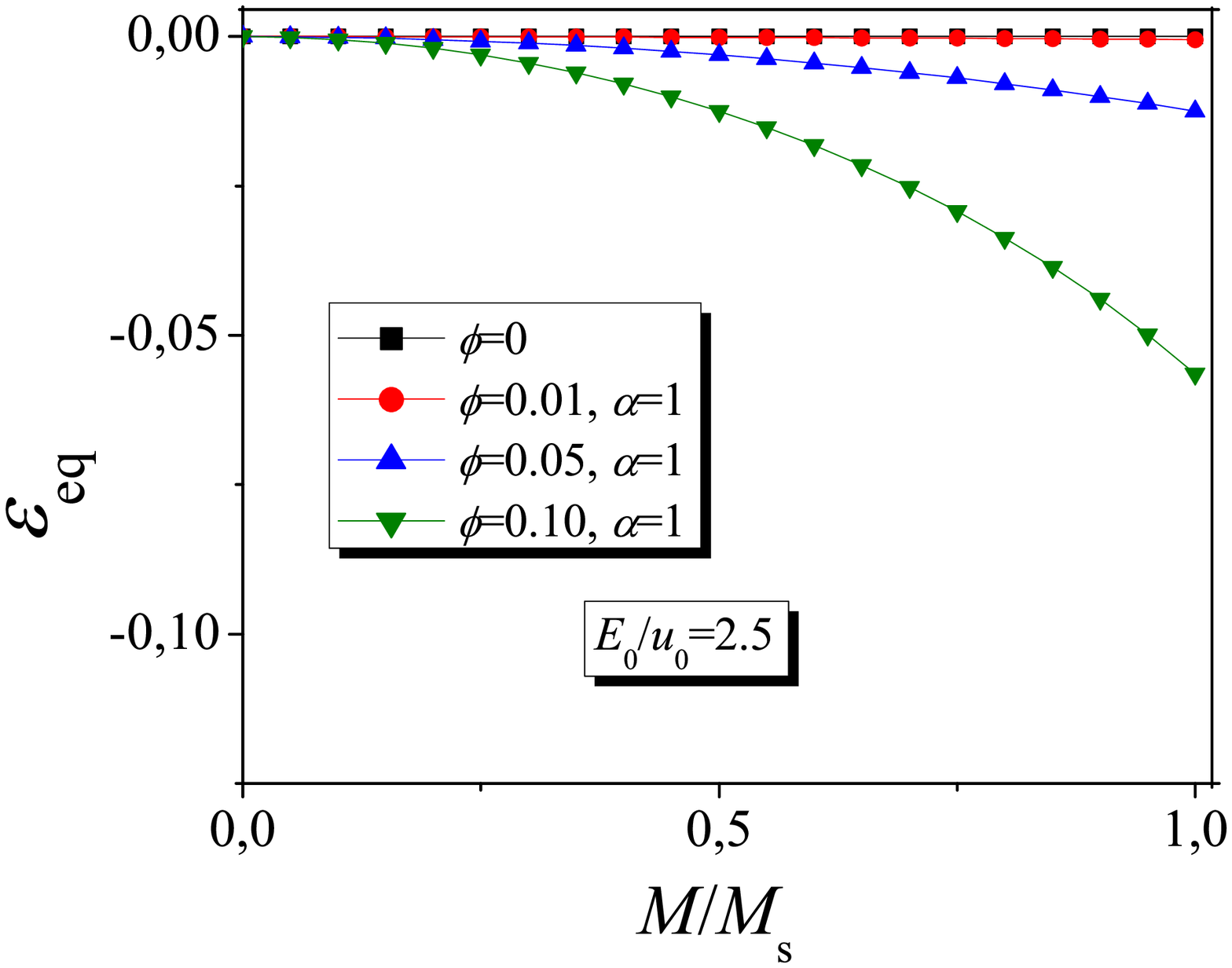}\\
(c) planes\\
\includegraphics[width=7cm]{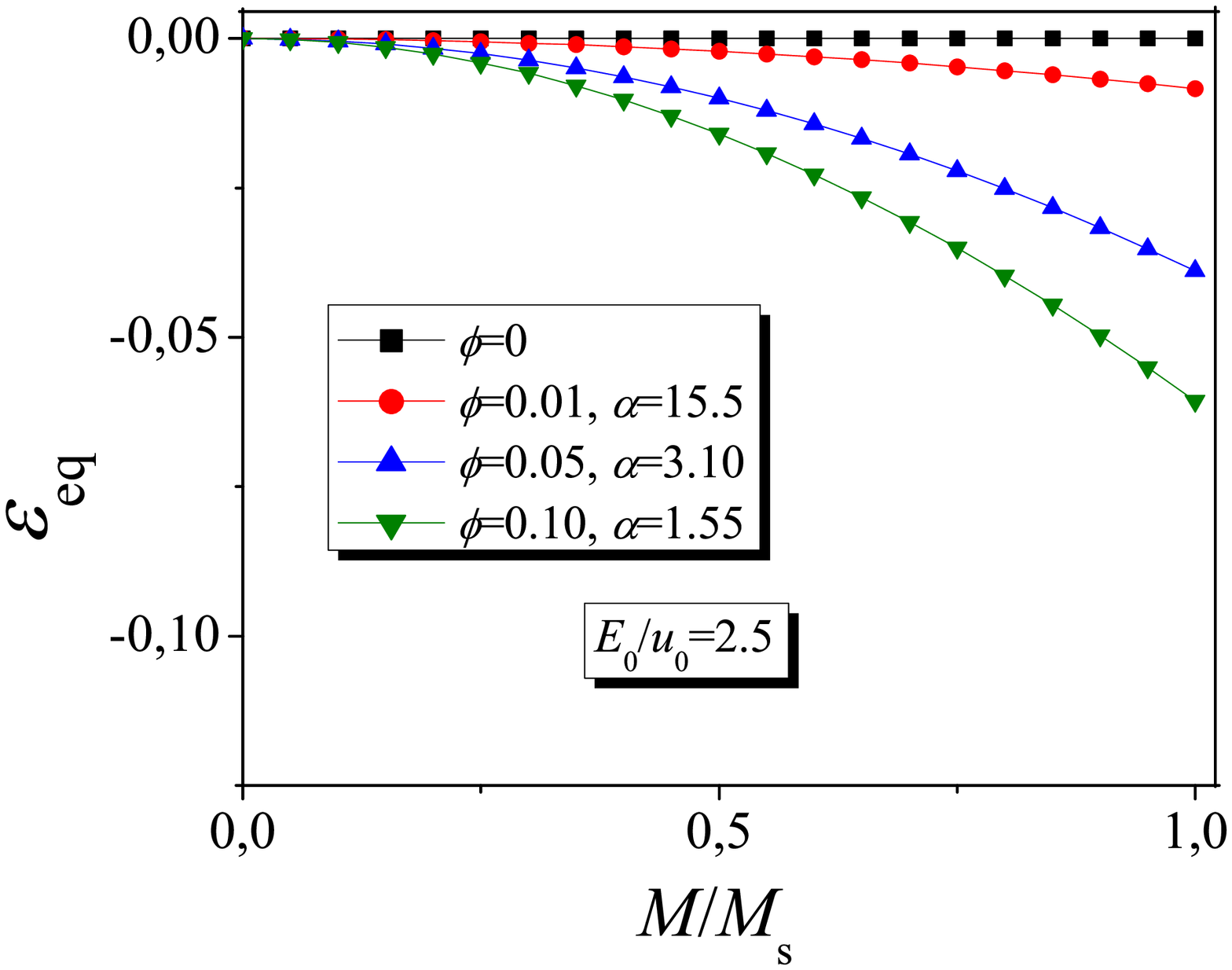}
\caption{\label{F5} Dependence of the equilibrium elongation $\varepsilon_{\rm eq}$ on the reduced magnetization $M/M_{\rm s}$ at different volume fractions $\phi$ and at fixed value of the parameter $E_0/u_0=2.5$. The values of the parameter $\alpha$ are given by: (a) Equation (\ref{ALP1}) for the chain-like distributions, (b) $\alpha=1$ for the isotropic distributions, (c) Equation (\ref{ALP2}) for the plane-like distributions.}
\end{figure}

\begin{figure}[!ht]
(a) chains\\
\includegraphics[width=7cm]{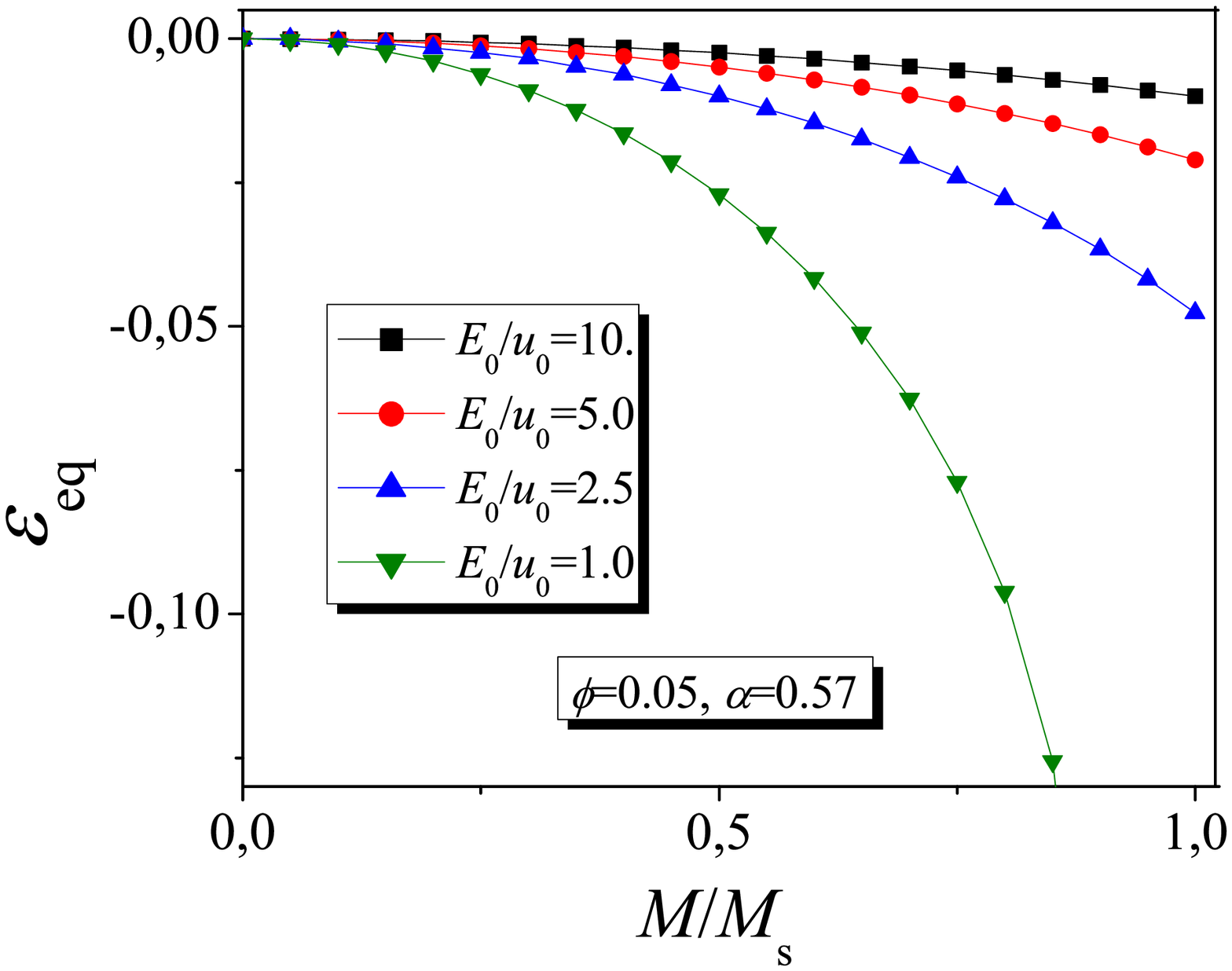}\\
(b) isotropic sample\\
\includegraphics[width=7cm]{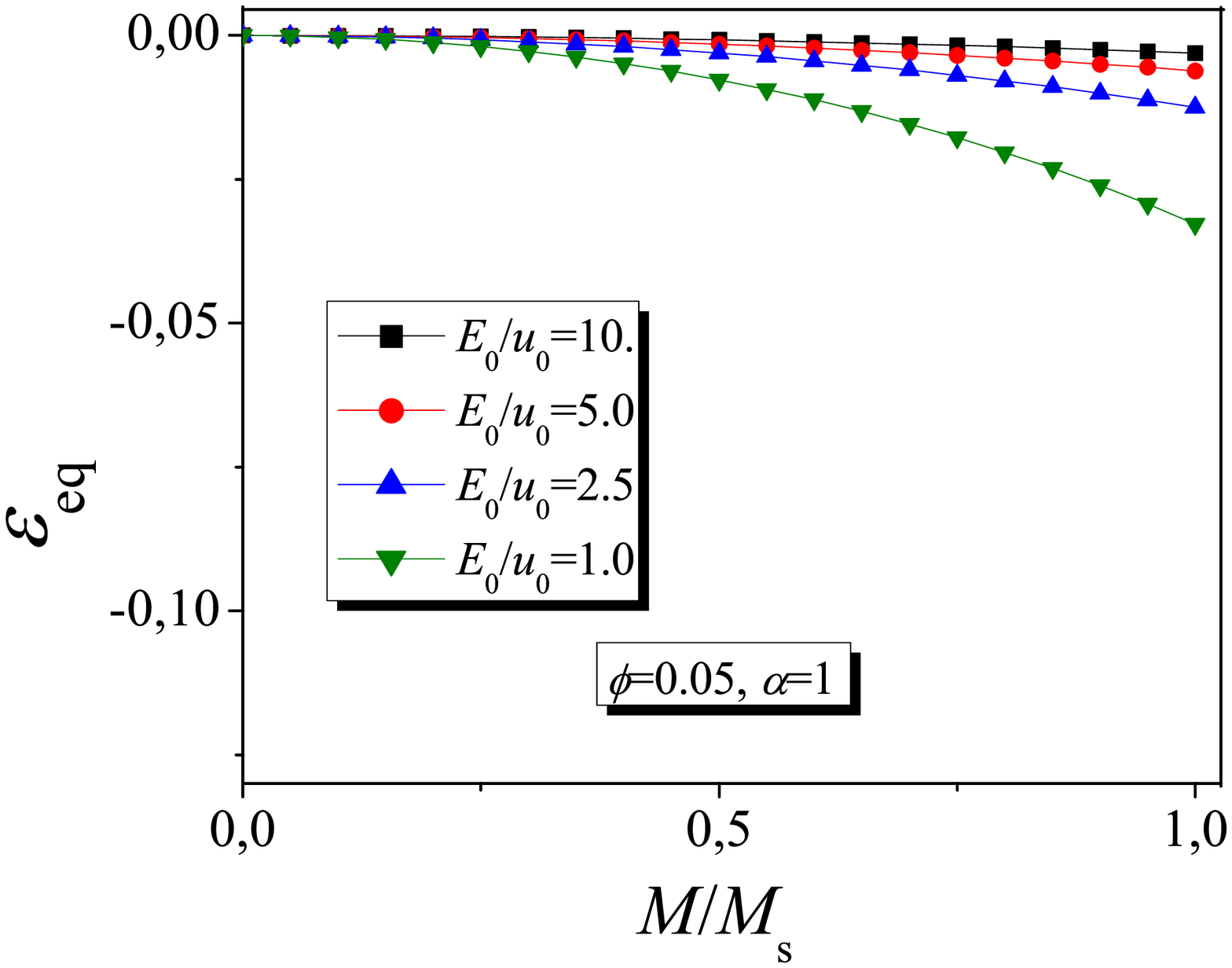}\\
(c) planes\\
\includegraphics[width=7cm]{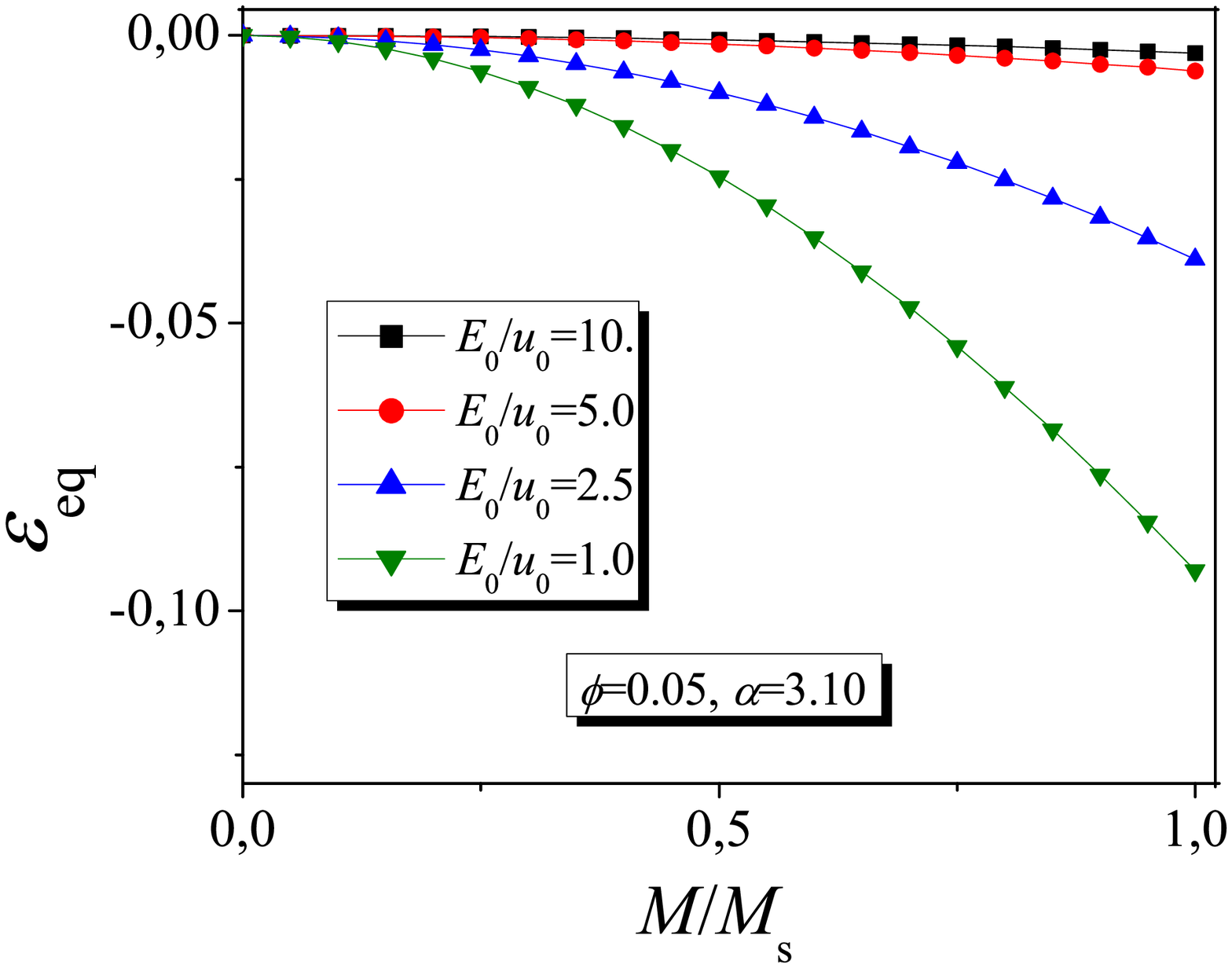}
\caption{\label{F6} Same as \ref{F5} but at different values of the parameter $E_0/u_0$ and at fixed volume fraction $\phi=0.05$.}
\end{figure}

We have solved Equation (\ref{EE}) numerically with respect to the parameter $\varepsilon_{\rm eq}$. \ref{F5} shows the dependence of the equilibrium elongation $\varepsilon_{\rm eq}$  on the reduced magnetization $M/M_{\rm s}$ for $E_0/u_0=2.5$ (that corresponds to $E_0=10^6~Pa$ and $u_0=4\times10^{5}~Pa$) and for different values of the volume fraction: $\phi=0$, $\phi=0.01$, $\phi=0.05$ and $\phi=0.1$. \ref{F6} shows the dependence of the equilibrium elongation $\varepsilon_{\rm eq}$ on the reduced magnetization $M/M_{\rm s}$ for $\phi=0.05$ and at different values of parameter $E_0/u_0$: $E_0/u_0=1.0$, $E_0/u_0=2.5$, $E_0/u_0=5.0$ and $E_0/u_0=10$. For each volume fraction  $\phi$ we have chosen the values of the structural parameter $\alpha$ given by Equation (\ref{ALP1}) for the chain-like distributions and given by Equation (\ref{ALP2}) for the plane-like distributions. We recall that Equation (\ref{ALP1}) and (\ref{ALP2}) describe such structures, in which the gaps between nearest particles are equal to the radius of particles. In the case of isotropic distribution we set $\alpha=1$ for any volume fraction $\phi$.

One can see from \ref{F5} and \ref{F6} that for any lattice structure a sample is uniaxially compressed along the direction of the external magnetic field, $\varepsilon_{\rm eq}<0$. With increasing value of $M/M_{\rm s}$ (i.e. with increasing magnetic field) the absolute value $|\varepsilon_{\rm eq}|$ increases. This means that the degree of uniaxial compression increases with increasing magnetic field. The sign of magnetostriction coincides with theoretical results obtained in Ref.$^[$\cite{Kankanala04,Martin06}$^]$ However, there exist some theoretical works, where the sign of magnetostriction differs from our result.$^[$\cite{Raikher00,Borcea01,Diguet10}$^]$ These works use the continuum mechanical approach and deal mainly with a homogeneous isotropic distribution of magnetic particles inside an MSE. The results of theoretical works$^[$\cite{Raikher00,Borcea01,Diguet10}$^]$ are in agreement with experiments which show that MSEs with homogeneous distribution of magnetic particles demonstrate a uniaxial expansion along the magnetic field.$^[$\cite{Zhou04,Diguet10}$^]$ On the other side, it was shown experimentally$^[$\cite{Jolly96a,Zhou04,Coquelle05}$^]$ that MSEs with the chain-like distributions of magnetic particles demonstrate a uniaxial compression along the magnetic field in agreement with our calculations.

\begin{figure}[!ht]
\includegraphics[width=6cm]{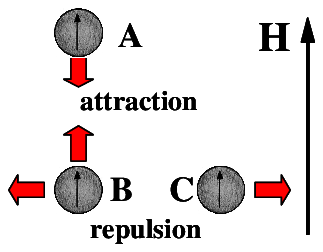}
\caption{\label{F7a} Attraction and repulsion of magnetic particles inside an MSE depending on their mutual positions.}
\end{figure}

One can expect that the mechanical behaviour of MSEs with the chain-like and plane-like distributions of particles are determined mainly by the attraction and repulsion of the particles as it is illustrated in \ref{F7a}. For the chain-like structures, the main contribution to the magnetic energy is due to the particles which lie ''in series'' to each other and attract to each other (A and B particles in \ref{F7a}), whereas for plane-like structures the main contribution to the magnetic energy is caused by the particles which lie ''in parallel'' to each other and repulse from each other (B and C particles in \ref{F7a}). In both configurations the total magnetic interaction leads to the contraction of a sample along the magnetic field in accordance with our results presented in \ref{F5} and \ref{F6}. Thus, we expect that the cubic lattice model is applicable to MSEs with the chain-like and plane-like distributions of magnetic particles, since this model takes explicitly into account the main interactions between magnetic particles in these structures (see \ref{F7a}). For homogeneous distribution, however, it is necessary to consider the effects of spatial distribution of particles on the mechanical behaviour of MSEs in more detail (including the calculation of the mechanical moduli) that can be a topic of further considerations.

Furthermore, one can see from \ref{F5} that the increase of the volume fraction $\phi$ results in the increase of the equilibrium elongation $|\varepsilon_{\rm eq}|$, when $M/M_{\rm s}$ is fixed. This is explained by the fact that the contribution of magnetic interaction becomes larger at higher values of $\phi$. One can see from \ref{F6} that the increase of the parameter $E_0/u_0$ results in the decrease of matrix deformation $|\varepsilon_{\rm eq}|$, when $M/M_{\rm s}$ is fixed. This is due to the fact that the relative contribution of magnetic interaction becomes smaller at larger values of the parameter $E_0/u_0$. Additionally, we can conclude from \ref{F6} that at fixed values of $M/M_{\rm s}$ and $\phi$ the magnitude of the deformation increases at decreasing $\alpha$ for $\alpha<1$ as well as at increasing $\alpha$ for $\alpha>1$. This result is explained as follows. At $\alpha<1$ the main contribution to the magnetic energy comes from the particles that lie in the same chains. With decreasing $\alpha$ (at $\alpha<1$) this contribution increases, since the distance between neighbouring particles in chains decreases and, as a result, the magnitude of $\varepsilon_{\rm eq}$ increases. On the other side, at $\alpha>1$ the main contribution to the magnetic energy comes from the particles that lie in the same planes. With increasing $\alpha$ (at $\alpha>1$) this contribution increases, since the distance between neighbouring particles in planes decreases and, as a result, the magnitude of $\varepsilon_{\rm eq}$ increases.

The next problem is to calculate mechanical moduli which characterize the response of MSEs to small external deformations. Note, that an MSE under magnetic field is an anisotropic medium. It is known that moduli for anisotropic media depend on the direction of small deformation with respect to the axis of anisotropy.$^[$\cite{Martinoty04,Stepanov07,Toshchevikov09,Toshchevikov10}$^]$ In the next sections we consider two types of small deformation applied to an MSE: shear deformation and  tensile deformation.

\subsection{Shear modulus of a magneto-sensitive elastomer}

\begin{figure}[!ht]
\includegraphics[width=8cm]{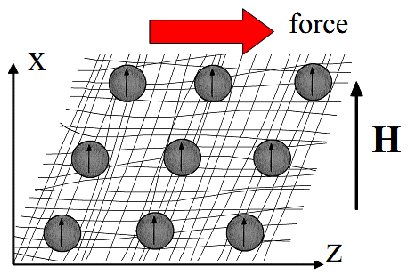}
\caption{\label{F7} Shear deformation of the MSE along the $z$-axis, perpendicular to the external magnetic field $\mathbf{H}$.}
\end{figure}

In the case of a shear deformation, we assume that the shear displacement is applied along the $z$-axis, i.e. it is perpendicular to the magnetic field; $\Delta (R_{ij})_z$ denotes the displacement of a particle in $z$ direction, see \ref{F7}. The shear strain is given by $\gamma=\Delta (R_{ij})_z/(R_{ij})_x$.

The new coordinates of the particles in the elastomer under both magnetic field and shear deformation are given by the following equations:
\begin{eqnarray}\label{RIJ1}
(R_{ij})_x &=& (R_{ij}^0)_x(1+\varepsilon_{\rm eq}),\nonumber\\
(R_{ij})_y &=& (R_{ij}^0)_y(1+\varepsilon_{\rm eq})^{-1/2},\\
(R_{ij})_z &=& (R_{ij}^0)_z(1+\varepsilon_{\rm eq})^{-1/2}+\gamma (R_{ij}^0)_x(1+\varepsilon_{\rm eq}).\nonumber
\end{eqnarray}
Here $(R_{ij}^0)_\xi$ are the components of vectors that separate $i$-th and $j$-th particle in the absence  of any fields.

The change of the free energy of an MSE, $\Delta F$, after small shear displacement $(\Delta R_{ij})_z$ from the equilibrium state (with $\varepsilon=\varepsilon_{\rm eq}$) can be written as:
\begin{equation}\label{FGD}
\Delta F=\frac{G_0\gamma^2}{2}+\Delta u(\gamma,\varepsilon_{\rm eq}),
\end{equation}
where $G_0=E_0/3$ is the shear modulus of a filled elastomer and $\Delta u(\gamma,\varepsilon_{\rm eq})$ is the change of the magnetic energy after the shear displacement from the equilibrium state:
\begin{equation}\label{FGDa}
\Delta u(\gamma,\varepsilon_{\rm eq})=u_2(\gamma,\varepsilon_{\rm eq})-u(\varepsilon_{\rm eq}).
\end{equation}
The value of $u(\varepsilon_{\rm eq})$ is given by Equation (\ref{UJ3}) and $u_2(\gamma,\varepsilon_{\rm eq})$ is determined by Equation (\ref{FE1}) and (\ref{UJ1}), in which, however, one should now substitute the values for $R_{ij}$ given by Equation (\ref{RIJ1}) for the shear deformation. Using Equation (\ref{FE1}), (\ref{UJ1}) and  (\ref{RIJ1}) we can rewrite Equation (\ref{FGD}) in the following form:
\begin{equation}\label{FGD1}
\Delta F=\frac{G_0\gamma^2}{2}+u_0 \phi^2\left(\frac{M}{M_{\rm s}}\right)^2 \left[f_2(\gamma,\varepsilon_{\rm eq})-f_2(0,\varepsilon_{\rm eq})\right],
\end{equation}
where the function $f_2(\gamma,\varepsilon_{\rm eq})$ has now the following form:
\begin{eqnarray} f_2(\gamma,\varepsilon_{\rm eq})=-\alpha(1+\varepsilon_{\rm eq})^{\frac{3}{2}}\sum\limits_{\{i_x i_y i_z\}\neq 0}\!\frac{(2-\gamma^2)\alpha^2i_x^2(1+\varepsilon_{\rm eq})^3-i_y^2-i_z^2-2\gamma\alpha i_x i_z(1+\varepsilon_{\rm eq})^{\frac{3}{2}}}{\left[(1+\gamma^2)\alpha^2i_x^2(1+\varepsilon_{\rm eq})^3+i_y^2+i_z^2+2\gamma\alpha i_x i_z(1+\varepsilon_{\rm eq})^{\frac{3}{2}}\right]^{\frac{5}{2}}}.
\end{eqnarray}

Shear modulus $G$ can be obtained as $G = \left(\partial^2 \Delta F/\partial \gamma^2\right)_{\gamma=0}$, that gives:
\begin{eqnarray}
G = G_0+u_0 \phi^2\left(\frac{M}{M_{\rm s}}\right)^2 3\alpha^3(1+\varepsilon_{\rm eq})^{\frac{9}{2}}\times \nonumber\\
\times\sum\limits_{\{i_x i_y i_z\}\neq 0}
\frac{i_x^2\left[4\alpha^4i_x^4(1+\varepsilon_{\rm
eq})^{6}+3\alpha^2i_x^2(i_y^2-9i_z^2)(1+\varepsilon_{\rm
eq})^{3}-i_y^4+3i_y^2i_z^2+4i_z^4\right]}
{\left[\alpha^2i_x^2(1+\varepsilon_{\rm
eq})^{3}+i_y^2+i_z^2\right]^{\frac{9}{2}}}.
\label{GEE}\end{eqnarray} We note that $G$ is an even function of
$M$, since it depends on $M$ through the factors $(M/M_{\rm s})^2$
and $\varepsilon_{\rm eq}$ that are both even functions of
$M$.$^[$\cite{Wood11}$^]$

\begin{figure}[!ht]
(a) chains\\
\includegraphics[width=7cm]{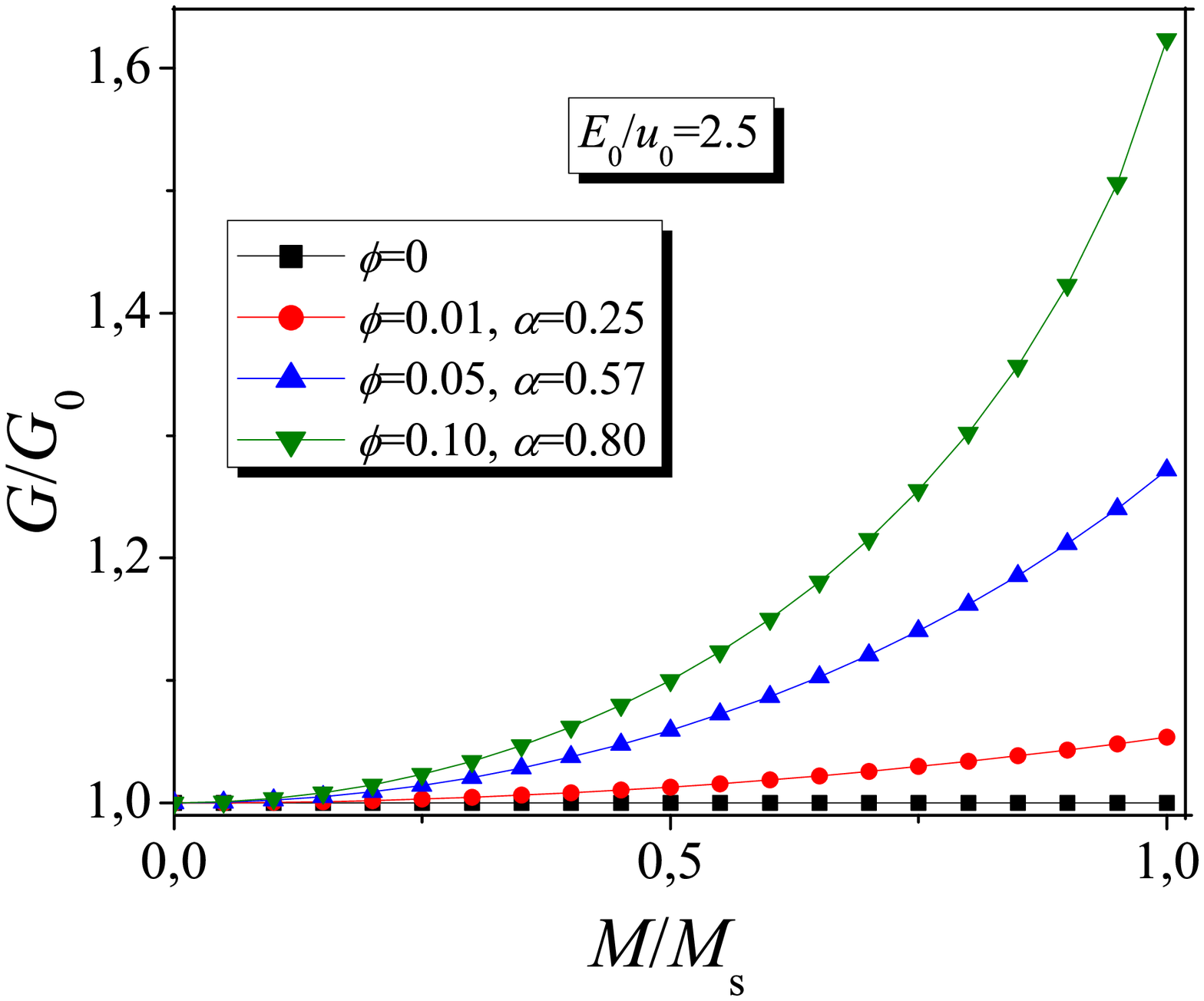}\\
(b) isotropic sample\\
\includegraphics[width=7cm]{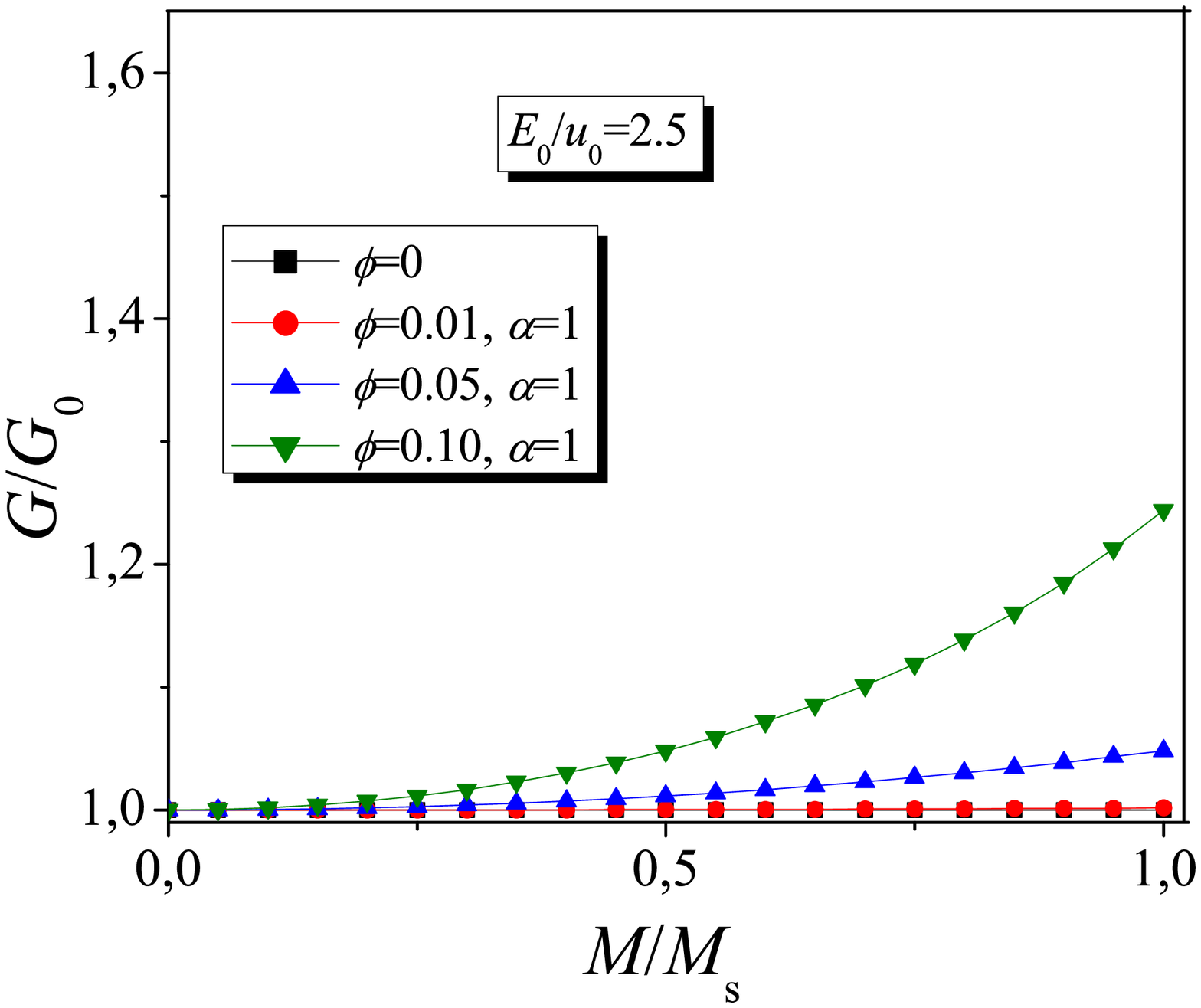}\\
(c) planes\\
\includegraphics[width=7cm]{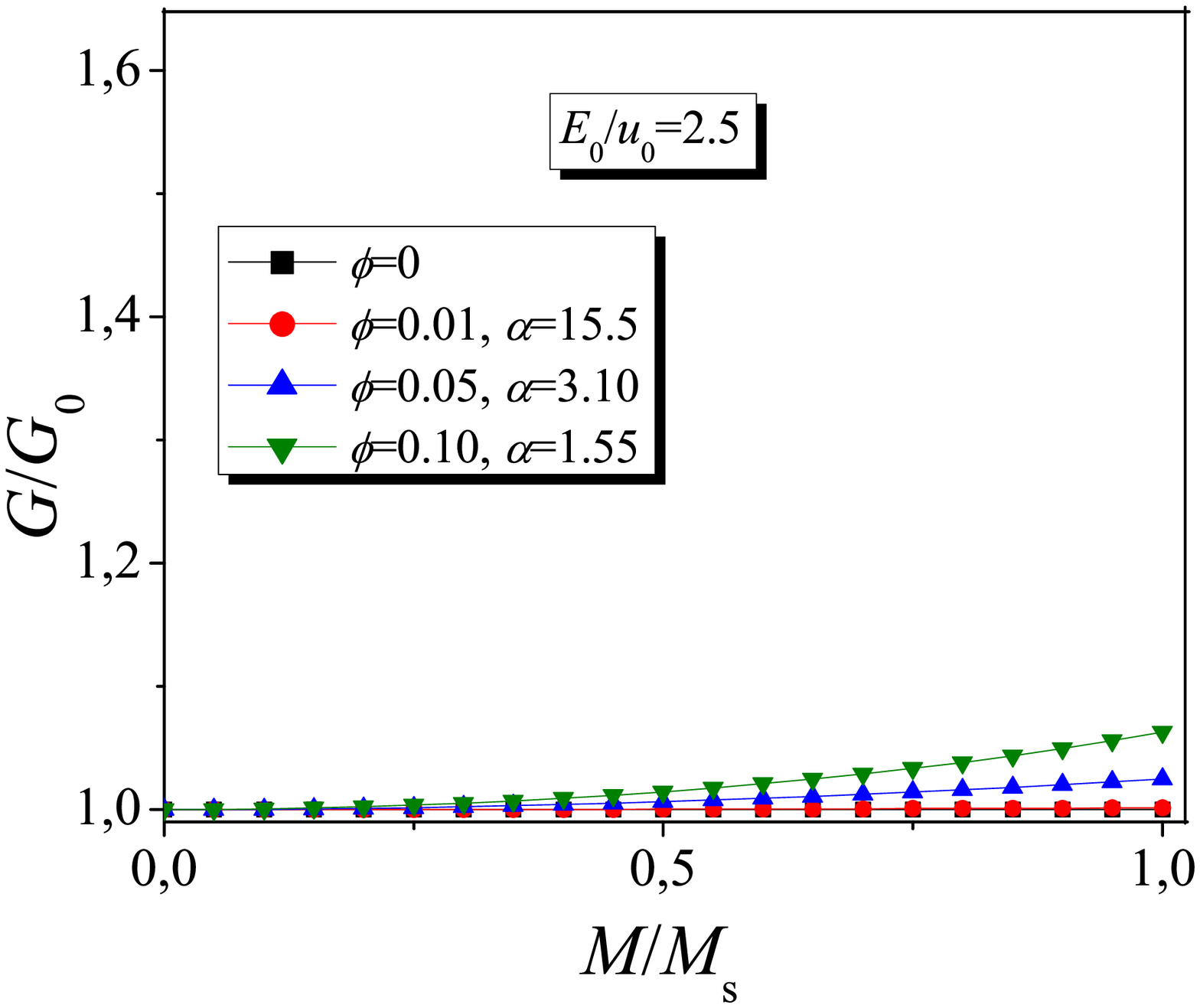}
\caption{\label{F8} Dependence of the shear modulus $G$ on the reduced magnetization $M/M_{\rm s}$ at different volume fractions $\phi$ and at fixed value of the parameter $E_0/u_0=2.5$. The values of the parameter $\alpha$ are given by: (a) Equation (\ref{ALP1}) for the chain-like distributions, (b) $\alpha=1$ for the isotropic distributions, (c) Equation (\ref{ALP2}) for the plane-like distributions.}
\end{figure}

\begin{figure}[!ht]
(a) chains\\
\includegraphics[width=7cm]{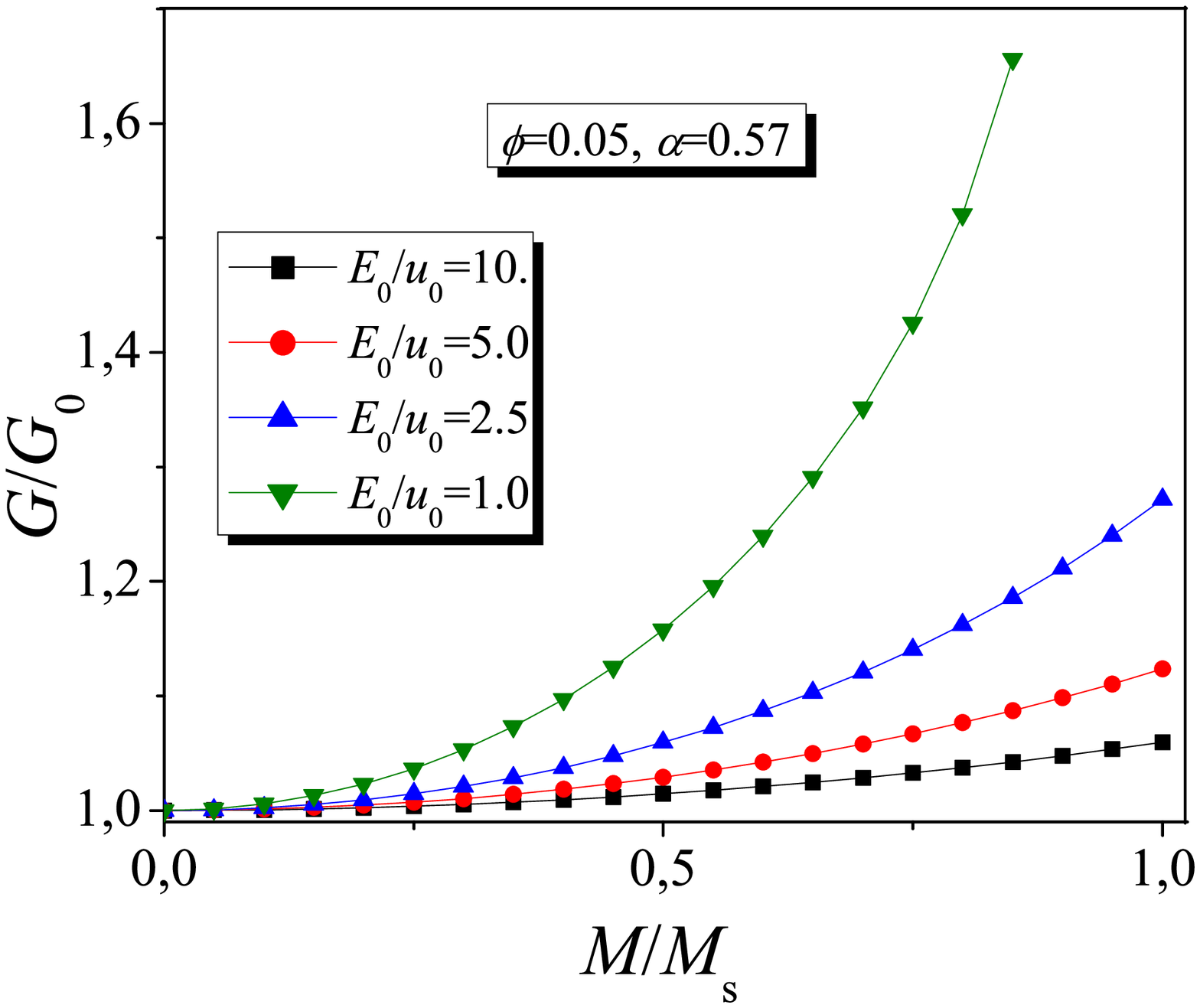}\\
(b) isotropic sample\\
\includegraphics[width=7cm]{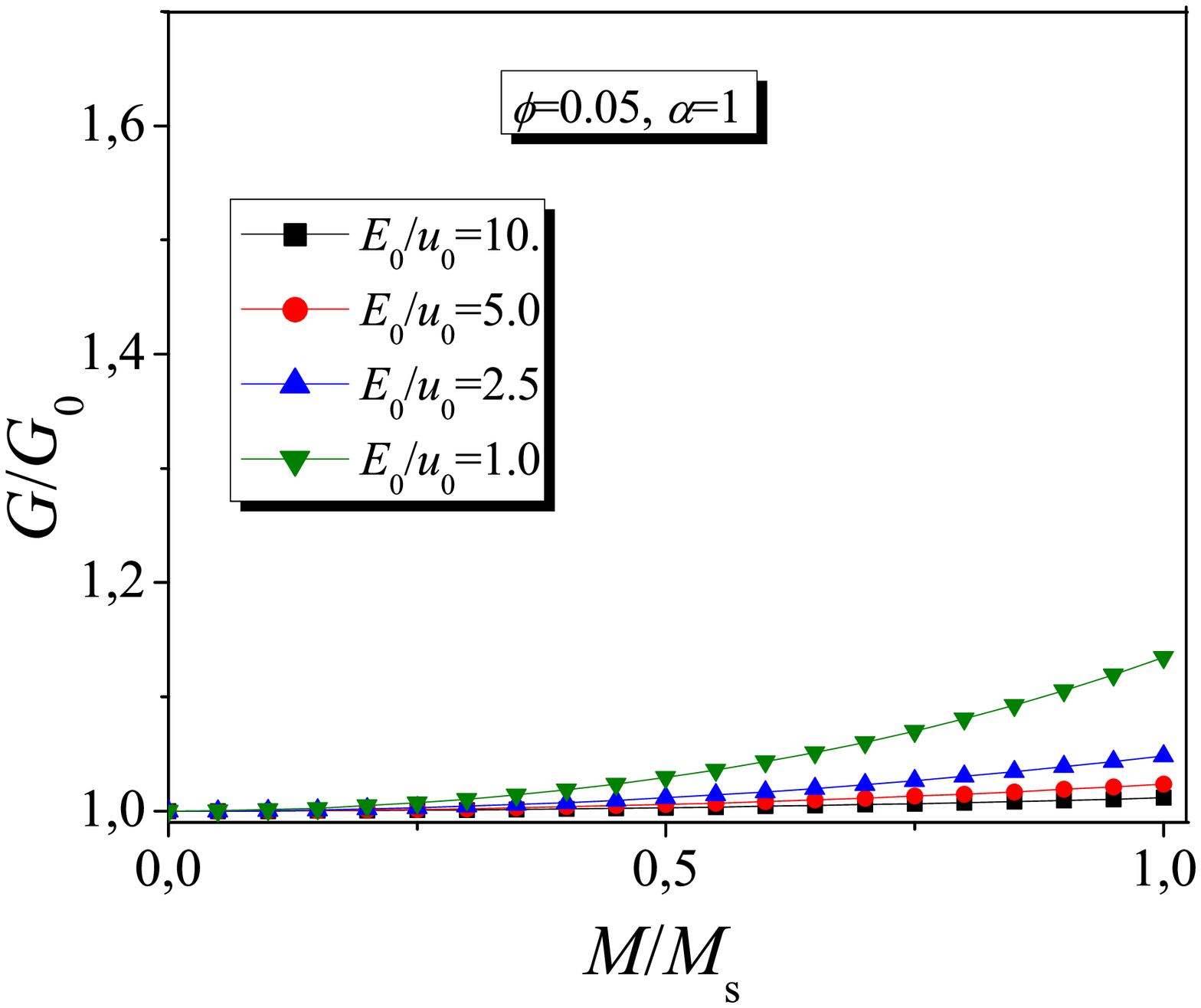}\\
(c) planes\\
\includegraphics[width=7cm]{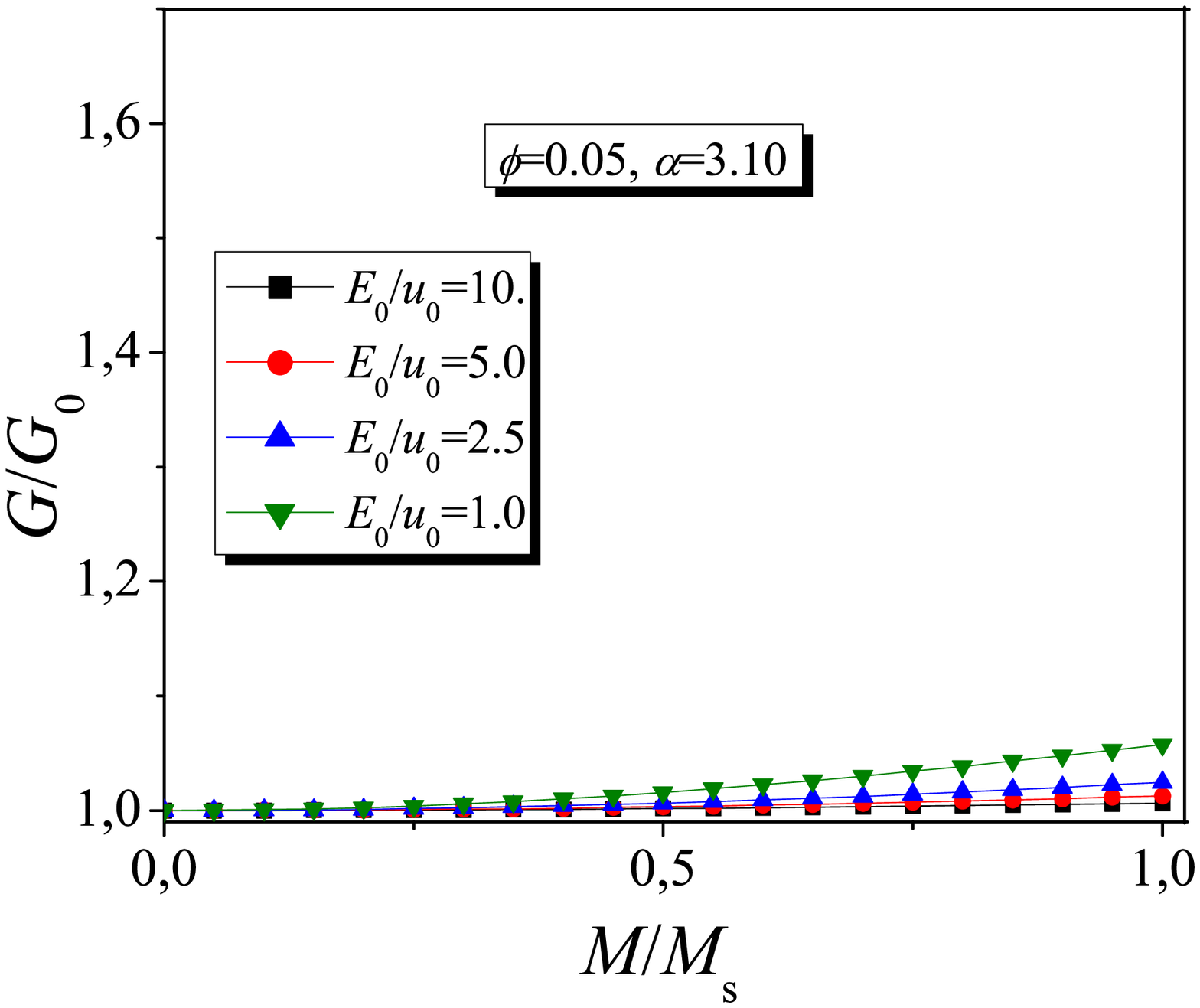}
\caption{\label{F9} Same as \ref{F8} but at different values of the parameter $E_0/u_0$ and at fixed volume fraction $\phi=0.05$.}
\end{figure}

Using Equation (\ref{GEE}), we have numerically  calculated $G$ as a function of the reduced magnetization $M/M_{\rm s}$ for different volume fractions $\phi$ and for different values of the parameter $E_0/u_0$.  We substituted into Equation (\ref{GEE}) the values of the equilibrium elongation $\varepsilon_{\rm eq}$, obtained from exact solution of Equation (\ref{EE}). Doing so, we take into consideration the effect of the magnetostriction on the shear modulus. As before, we have chosen the following values of the parameter $\alpha$: for the chain-like and plane-like structures of the particles, the values of $\alpha$ are given by Equation (\ref{ALP1}) and (\ref{ALP2}), respectively; for the isotropic distribution of particles we set $\alpha=1$. \ref{F8} shows the dependence of the shear modulus $G$ on the reduced magnetization $M/M_{\rm s}$ at different values of the volume fraction $\phi$: $\phi=0$, $\phi=0.01$, $\phi=0.05$, $\phi=0.1$ and at the fixed value of the parameter $E_0/u_0=2.5$ (e.g. $E_0=10^6~Pa$, $u_0=4\times10^{5}~Pa$). \ref{F9} shows the dependence of the shear modulus $G$ on the reduced magnetization $M/M_{\rm s}$ at different values of the  parameter $E_0/u_0$: $E_0/u_0=1.0$, $E_0/u_0=2.5$, $E_0/u_0=5.0$, $E_0/u_0=10$ and at fixed value of the volume fraction $\phi=0.05$.

One can see from \ref{F8} and \ref{F9} that the shear modulus $G$ increases at increasing magnetization for all distributions of magnetic particles. As in Refs.$^[$\cite{Jolly96,Jolly96a,Zhu06}$^]$ this effect is due to the fact that under shearing of an MSE an additional force of magnetic interaction between particles appears, which increases the values of the modulus. Moreover, as it can be seen from \ref{F8} and \ref{F9}, the value of  $G$ increases only slightly for the isotropic and plane-like distributions of magnetic particles, as compared with the chain-like distributions. This can be explained by especially strong magnetic interactions between particles in the chain-like structures. The total force of these pair-wise interactions is directed along the axis of chain, which makes this type of structure strongly resistant against the shearing perpendicular to the chains. Substituting in Equation (\ref{GEE}) values $i_y=0$ and $i_z=0$ we recover the result obtained by Jolly et al.,$^[$\cite{Jolly96,Jolly96a}$^]$ who considered only a one-chain structure. Comparing the multi-chain result with the one-chain result one can see that effect of the neighbouring chains in a multi-chain system reduces the change of the modulus, $G-G_0$, as compared to the value $G-G_0$ for a one-chain system.

From \ref{F8} one can see that the increase of volume fraction $\phi$ leads to the increase of the shear modulus $G$ at fixed $M/M_{\rm s}$. This is explained by the fact that the relative contribution of magnetic interaction becomes larger at higher values of $\phi$. From \ref{F9} it follows that the increase of the parameter $E_0/u_0$ results in the decrease of the shear modulus $G$, when $M/M_{\rm s}$ is fixed. This is due to the fact that the relative contribution of magnetic interaction becomes smaller at higher values of the parameter $E_0/u_0$. Additionally, we can conclude from \ref{F9} that the relative change $G/G_0$ increases at decreasing $\alpha$ and at fixed values of $M/M_{\rm s}$ and $\phi$. This tendency is explained by the fact that the shear modulus is determined by the magnetic interactions between the particles, that are shifted at the shear deformation (i.e. that lie along the $x$-axis). With decreasing $\alpha$ the interaction between these particles increases, since the distance between neighbouring particles along the $x$-axis decreases and, as a result, the value $G/G_0$ increases, see \ref{F9}.

\subsection{Young's  (tensile) modulus of a magneto-sensitive elastomer}

\begin{figure}[!ht]
\includegraphics[width=8cm]{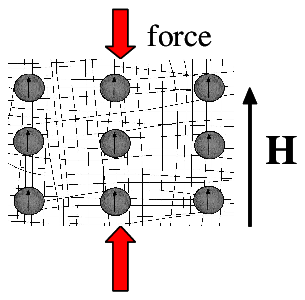}
\caption{\label{F10} Tensile deformation of an MSE along the external magnetic field $\mathbf{H}$.}
\end{figure}

In the case of a tensile deformation, we consider such geometry, when an additional small mechanical force is applied along the external magnetic field $H$, as it is shown in \ref{F10}.

As it follows from Equation (\ref{FE2}), (\ref{UJ3}) and  (\ref{FFF}), the free energy of an MSE as a function of $\varepsilon$  has the following form:
\begin{equation}\label{FED}
F=\frac{E_0\varepsilon^2}{2}+u_0 \phi^2\left(\frac{M}{M_{\rm s}}\right)^2 f(\varepsilon),
\end{equation}
where $f(\varepsilon)$ is given by Equation (\ref{FFF}). Note, that $\varepsilon$ in Equation (\ref{FED}) is the total strain which includes both $\varepsilon_{\rm eq}$ and additional small deformation. Thus, the Young's  modulus $E$ for an elastomer compressed by the magnetic field until the relative deformation $\varepsilon_{\rm eq}$ can be obtained as the second derivative of the free energy with respect to $\varepsilon$: $E = \left(\partial^2 F/\partial \varepsilon^2\right)_{\varepsilon=\varepsilon_{\rm eq}}$  that gives:

\begin{eqnarray}
E=E_0-u_0 \phi^2\left(\frac{M}{M_{\rm s}}\right)^2\frac{3\alpha}{4\sqrt{1+\varepsilon_{\rm eq}}}\times\nonumber\\
\times\sum\limits_{\{i_x i_y i_z\}\neq 0} \frac{32\alpha^6i_x^6(1+\varepsilon_{\rm eq})^9-192\alpha^4i_x^4(i_y^2+i_z^2)(1+\varepsilon_{\rm eq})^6+90\alpha^2i_x^2(i_y^2+i_z^2)^2(1+\varepsilon_{\rm eq})^3-(i_y^2+i_z^2)^3}{\left[\alpha^2i_x^2(1+\varepsilon_{\rm eq})^{3}+i_y^2+i_z^2\right]^{\frac{9}{2}}}.\label{EEEE}
\end{eqnarray}
We note that $E$ is an even function of $M$, since it depends on
$M$ through the factors $(M/M_{\rm s})^2$ and $\varepsilon_{\rm
eq}$ that are both even functions of $M$.

\begin{figure}[!ht]
(a) chains\\
\includegraphics[width=7cm]{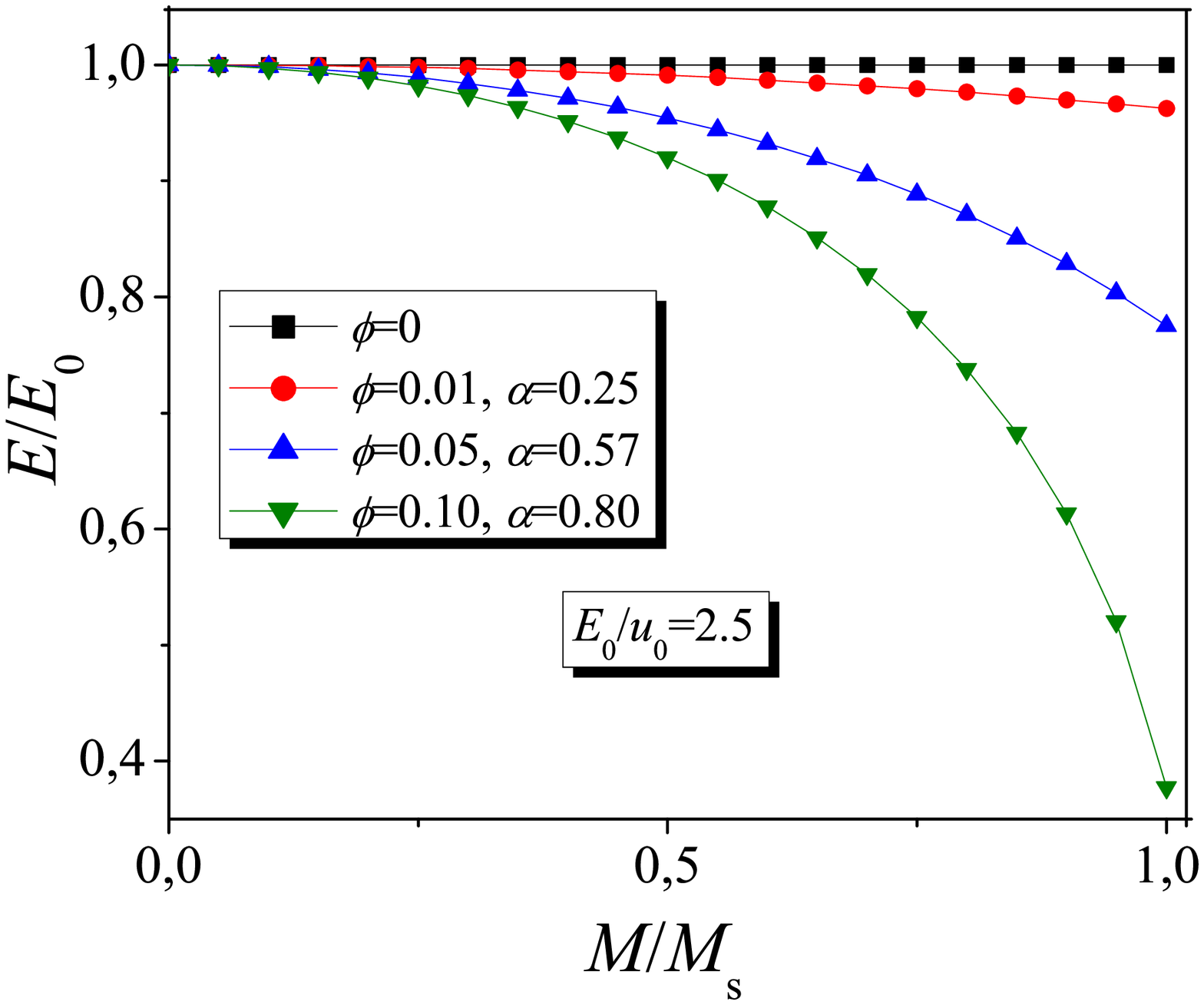}\\
(b) isotropic sample\\
\includegraphics[width=7cm]{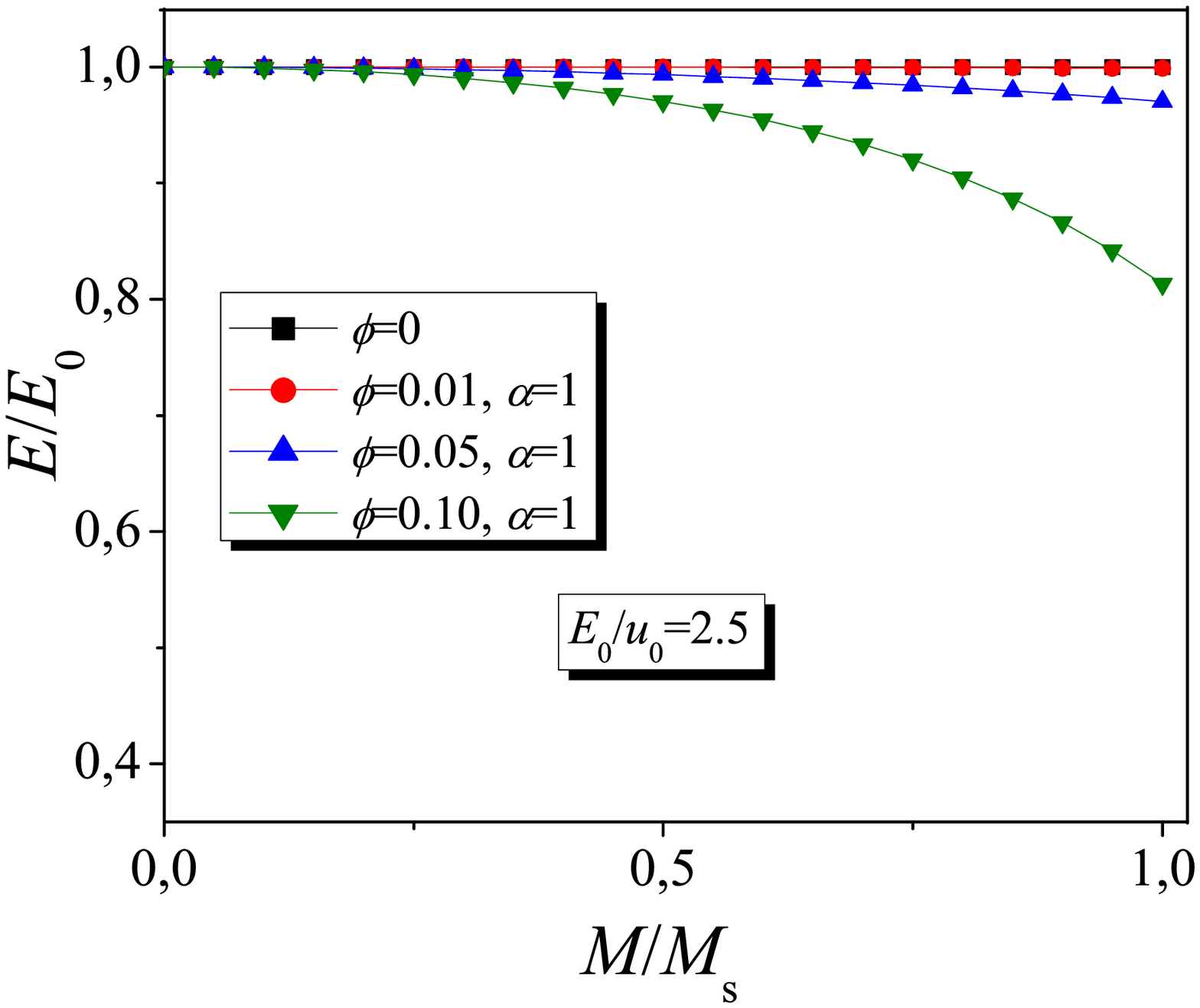}\\
(c) planes\\
\includegraphics[width=7cm]{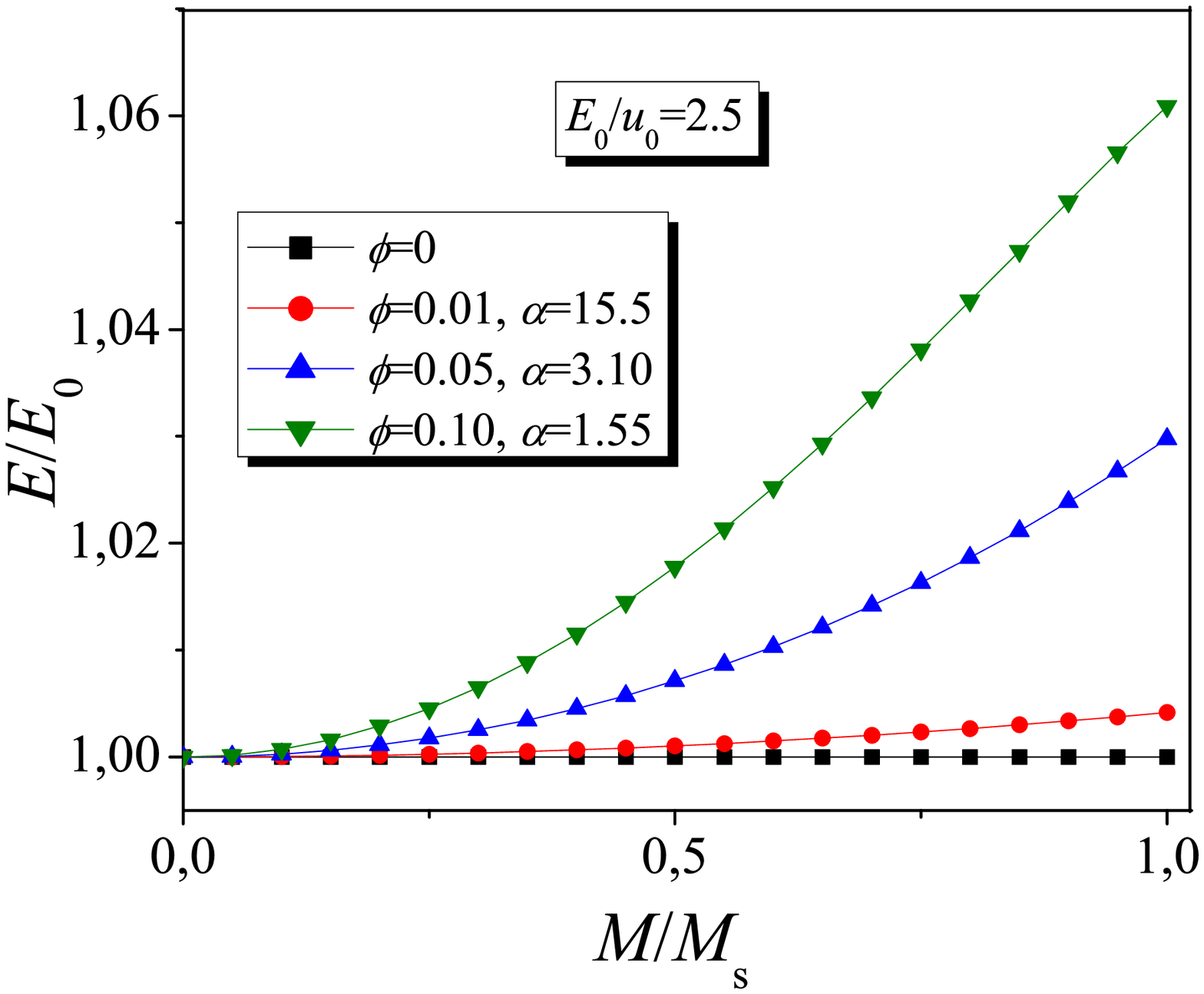}
\caption{\label{F11} Dependence of the Young's  modulus $E$ on the reduced magnetization $M/M_{\rm s}$ at different volume fractions $\phi$ and at fixed value of the parameter $E_0/u_0=2.5$. The values of the  parameter $\alpha$ are given by: (a) Equation (\ref{ALP1}) for the chain-like distributions, (b) $\alpha=1$ for the isotropic distributions, (c) Equation (\ref{ALP2}) for the plane-like distributions.}
\end{figure}

\begin{figure}[!ht]
(a) chains\\
\includegraphics[width=7cm]{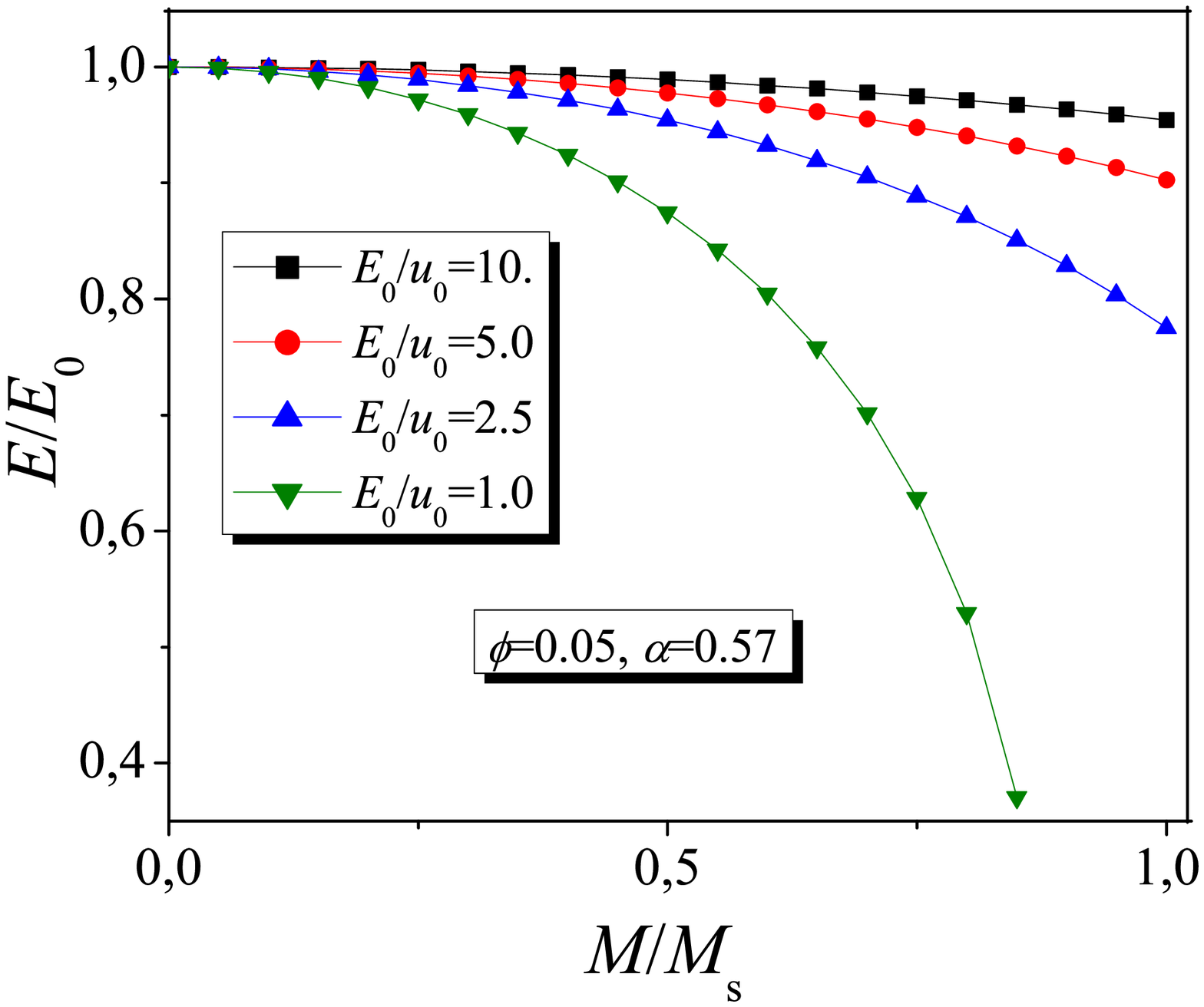}\\
(b) isotropic sample\\
\includegraphics[width=7cm]{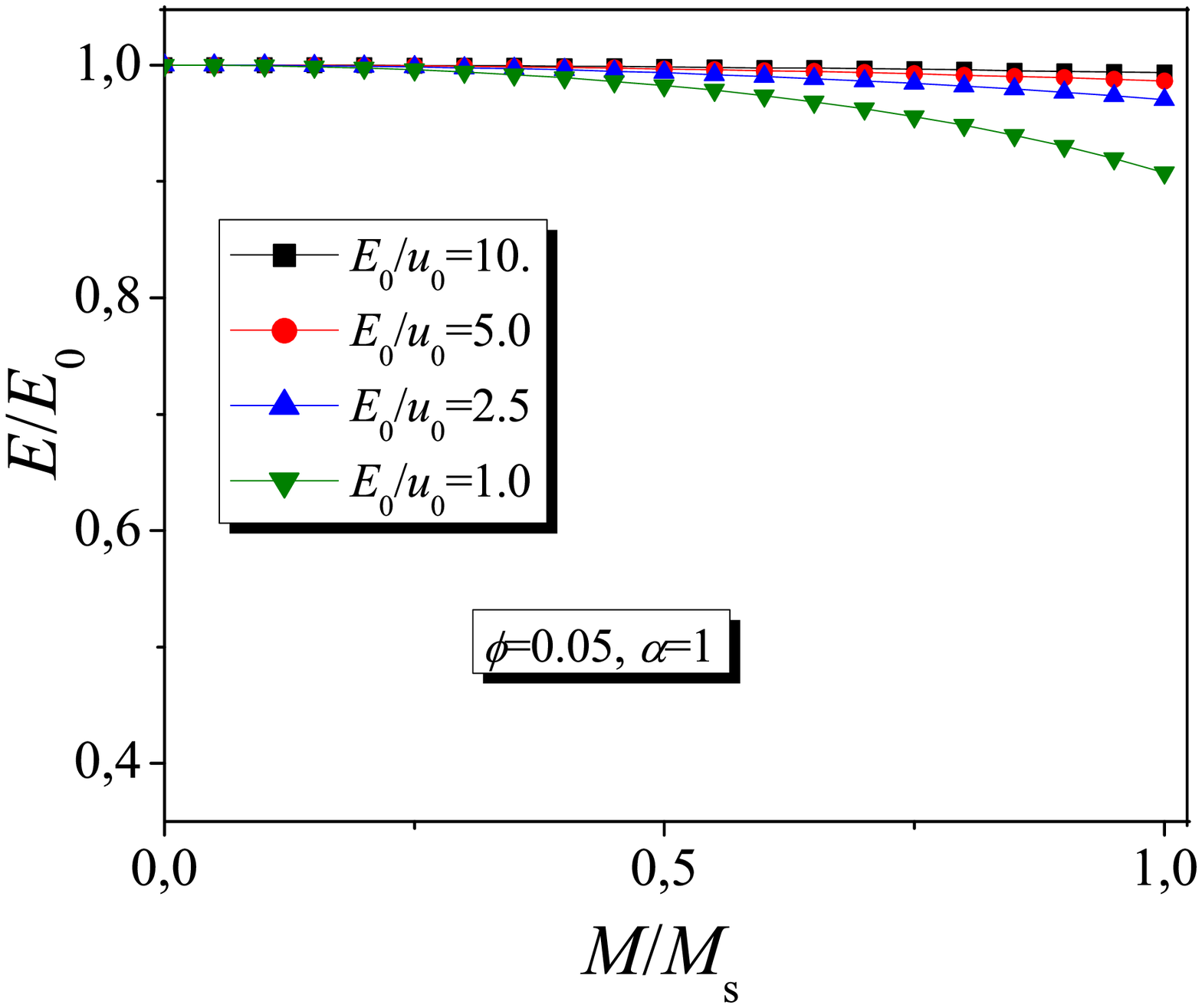}\\
(c) planes\\
\includegraphics[width=7cm]{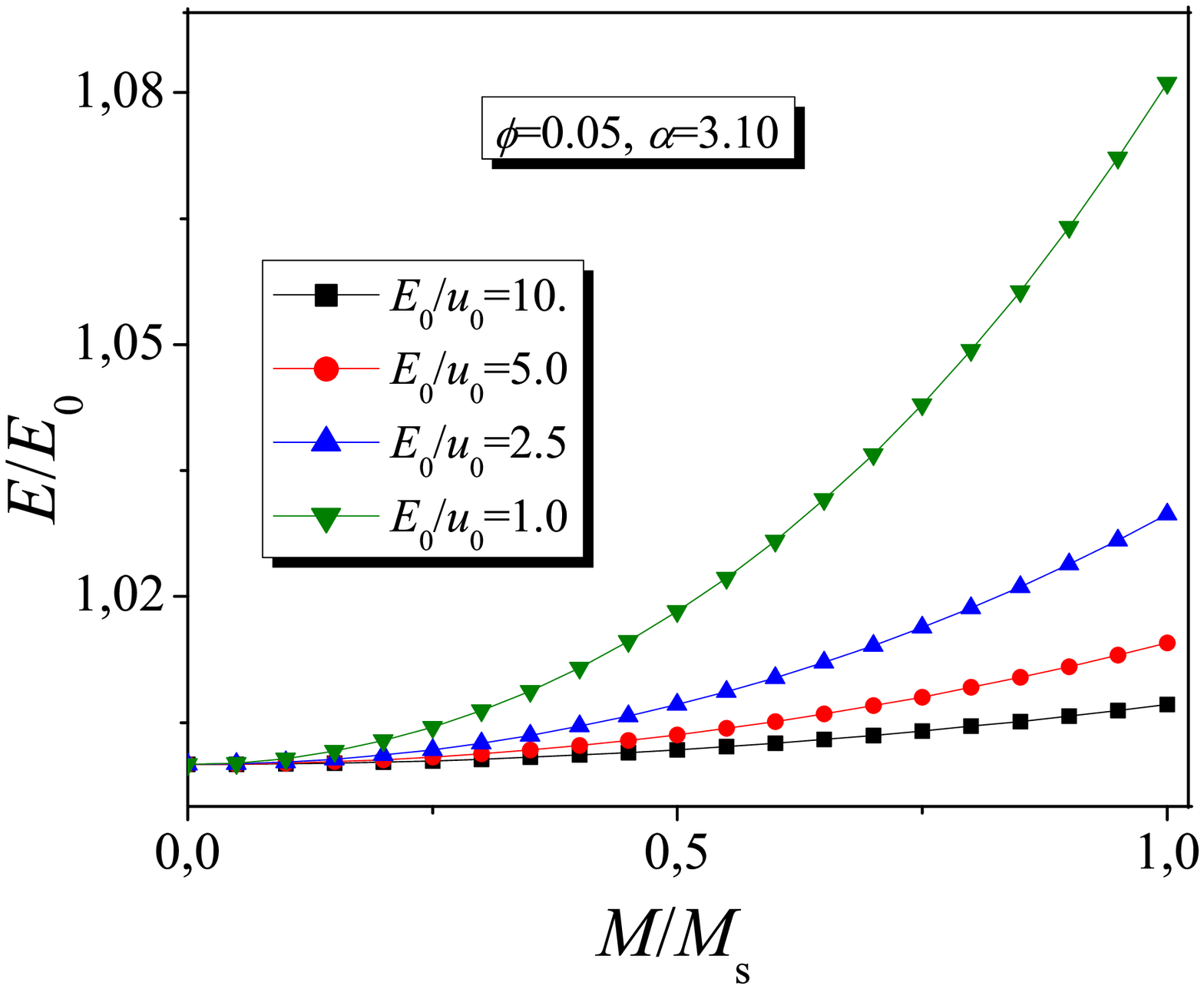}
\caption{\label{F12} Same as \ref{F11} but at different values of the parameter $E_0/u_0$ and at fixed volume fraction $\phi=0.05$.}
\end{figure}

Using Equation (\ref{EEEE}) we have numerically calculated $E$ as a function of the reduced magnetization $M/M_{\rm s}$ at varied values of the parameters $\phi$ and $E_0/u_0$. The results are presented in \ref{F11} and \ref{F12}. As in the previous section, we have substituted into Equation (\ref{EEEE}) the values of the equilibrium elongation $\varepsilon_{\rm eq}$ obtained from exact solution of Equation (\ref{EE}) and the parameter $\alpha$ is chosen according to Equation (\ref{ALP1}) and (\ref{ALP2}) for the chain-like and plane-like distributions of the magnetic particles, respectively. \ref{F11} shows the dependence of the Young's  modulus $E$ on the reduced magnetization $M/M_{\rm s}$ at different values of the volume fraction  $\phi$: $\phi=0$, $\phi=0.01$, $\phi=0.05$, $\phi=0.1$ and at fixed value of the parameter $E_0/u_0=2.5$. \ref{F12} shows the  dependence of the Young's  modulus $E$ on the reduced magnetization $M/M_{\rm s}$ at different values of the parameter $E_0/u_0$: $E_0/u_0=1.0$, $E_0/u_0=2.5$, $E_0/u_0=5.0$, $E_0/u_0=10$ and at the fixed volume fraction $\phi=0.05$.

\begin{figure}[!ht]
\includegraphics[width=14cm]{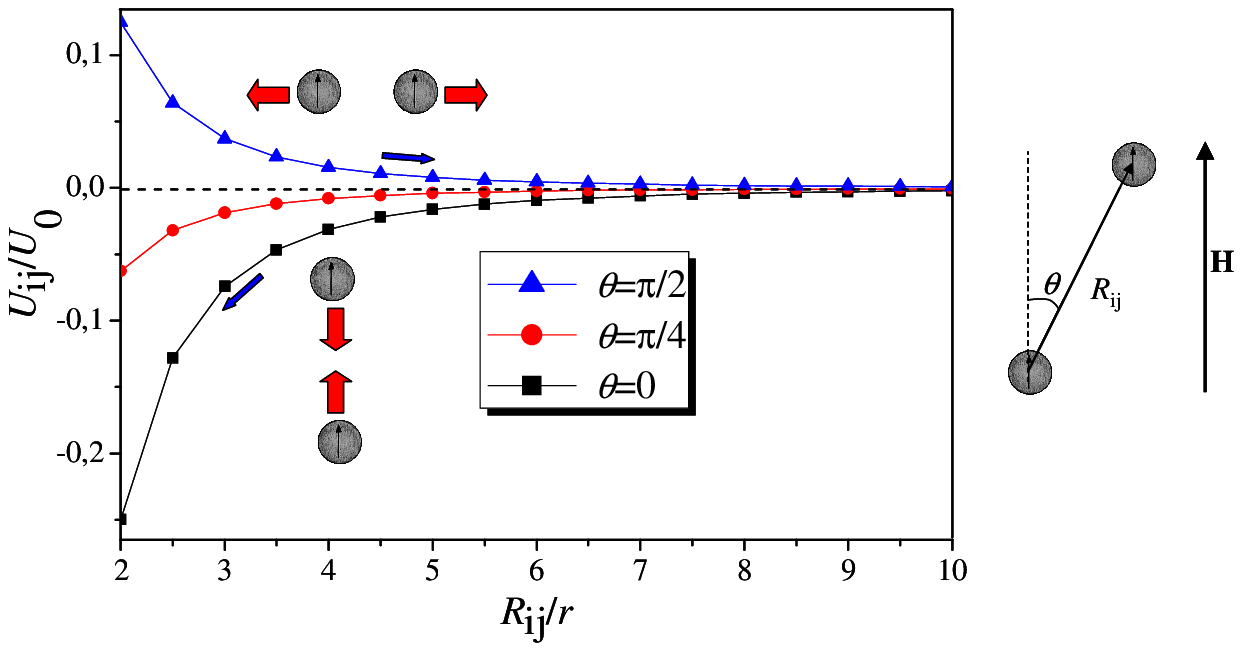}
\caption{\label{F14} Reduced interaction energy, $U_{ij}/U_0$, between two point-like magnetic dipoles as a function of the reduced distance $R_{ij}/r$ between the magnetic particles and at different orientation angles, $\theta$.  Here $U_0=\mu_0m^2/4\pi r^3$, where $m$ and $r$ are the magnetic moment and the radius of magnetic particles, respectively.}
\end{figure}

One can see from \ref{F11} and \ref{F12} that with increasing magnetization (i.e. with increasing magnetic field) the Young's  modulus $E$ decreases for the chain-like and isotropic distributions of magnetic particles and increases for the plane-like distributions. This is because in the chain-like structures of magnetic particles the main contribution to the magnetic energy comes from the interactions between particles in a chain. The potential of such interactions has a negative sign and goes to $-\infty$, when the distance between particles goes to 0, see \ref{F14} for $\theta=0$. Increase of the magnetic field leads to a greater attractive force between neighbouring particles and, thus, the contraction of the chain is energetically favourable. Moreover, in this case the curvature of the magnetic potential as a function of the distance between particles is negative and decreases with increasing magnetic field. This leads to the decrease of the modulus $E$ of the MSE  with increasing magnetic field. The opposite situation takes place in the plane-like structures of the magnetic particles. The main contribution to the magnetic energy comes from the interactions of particles in planes, where the potential of such interactions has a positive sign and goes to $+\infty$, when the distance between particles goes to 0, see \ref{F14} for $\theta=\pi/2$. Increase of the magnetic field leads to the situation when magnetic particles repulse, because it is energetically favourable. In this case the curvature of the magnetic potential is positive and increases with increasing magnetic field. This leads to an increase of the modulus $E$ of the MSE with increasing magnetic field. It turns out that for isotropic distribution of particles inside an MSE the main contribution  to the magnetic energy comes from the particles, which lie ''in line'' to each other (for $\theta=0$ in \ref{F14}). Therefore, for the isotropic distribution of particles, the modulus $E$ decreases under magnetic field as for the chain-like distribution of particles.

Furthermore, it can be seen from \ref{F11} that the increase of  the volume fraction  $\phi$  leads to the increase of the absolute values of the change  of the modulus $|E-E_0|$ for all distributions at fixed $M/M_{\rm s}$. From \ref{F12} it follows that the increase of the parameter $E_0/u_0$ results in the decrease of the absolute values $|E-E_0|$ for all distributions at fixed $M/M_{\rm s}$. These results are explained by the facts that the relative contribution of the magnetic energy to the modulus increases at increasing values of the volume fraction $\phi$ and decreases at increasing values of the parameter $E_0/u_0$. Additionally, we can conclude from \ref{F12} that the value $E/E_0$ increases at increasing $\alpha$ and at fixed values of $M/M_{\rm s}$ and $\phi$. This tendency is explained as follows. At increasing $\alpha$ the curvatures of the functions $U_{ij}$ increase both for the particles that lie along the $x$-axis (since $L_x^{(0)}$ increases) and for the particles that lie along the $y$- and $z$-axes (since $L_y^{(0)}$ and $L_z^{(0)}$ decrease), see \ref{F14}. Both these effects lead to the increase of the value $E/E_0$ at increasing $\alpha$, see \ref{F12}, since the curvature of $U_{ij}$ is proportional to the Young's modulus, $E = \partial^2 F/\partial \varepsilon^2$.

\section{Discussion} \label{S4a}
In this section we would like to compare some of our findings with predictions of other theories as well as with existing experimental data. First of all, we should mention that in our studies we have used a lattice model to describe the distribution of magnetic particles in a magneto-sensitive elastomer. The simple cubic lattice allowed us to consider different particle distributions including the chain-like, isotropic and plane-like distribution. In all cases we obtained the negative sign of magnetostriction effect, i.e. the sample slightly contracts under application of a homogeneous magnetic field. The predicted magnitude of deformation for chain-like structures does not exceed 5\% at the highest strength of magnetic field, which is in a quantitative agreement with experimental data.$^[$\cite{Jolly96a,Coquelle05}$^]$ In the case of isotropic distribution our result of equilibrium contraction disagrees with experimental results, where the sample elongation less than 1\% has been observed.$^[$\cite{Bellan02}$^]$ We suppose that this discrepancy between the theory prediction and the experimental finding arises from the fact that the ''isotropic'' distribution on a cubic lattice does not correspond to a distribution of particles in a real isotropic composite. A better approximation to the real isotropic distribution would be the volume-centered lattice packing, which will be the topic of future studies.

There exist theoretical studies predicting the positive sign of magnetostriction effect in isotropic composites which are considered as a continuous medium.$^[$\cite{Borcea01,Raikher03,Diguet10}$^]$ It is known from the electrodynamics that a homogeneous magnetic sphere, brought into a homogeneous magnetic field, elongates along the field.$^[$\cite{Landau80}$^]$ This effect is indeed observed in ''ferrogels'',$^[$\cite{Zrinyi97,Zrinyi98,Varga05,Varga05a,Gollwitzer08,Filipcsei10}$^]$ where magnetic particles can diffuse through the mesh of the matrix and build elongated clusters under application of external magnetic field. Due to this effect the ''ferrogel'' sample exhibits a macroscopic elongation. In the case of MSEs we can not use the theory of continuous medium, because the micro-sized magnetic particles cannot diffuse through the mesh of polymer network and rearrange their mutual positions.

\begin{figure}[!ht]
\includegraphics[width=5cm]{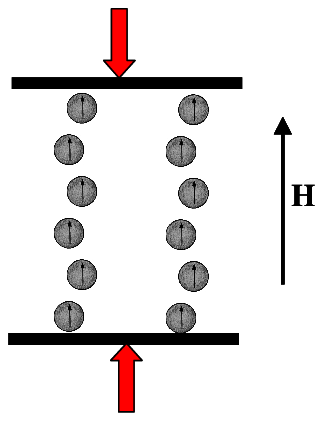}
\caption{\label{F15}  Tensile deformation of an MSE with ''wave-like'' irregular chains of magnetic particles along the external magnetic field $\mathbf{H}$.}
\end{figure}

Another question concerns the decrease of the Young's modulus predicted by our theory for the case of chain-like particle distributions. Some experiments show an opposite tendency, i.e. the Young's modulus increases under application of the magnetic field.$^[$\cite{Bellan02,Varga06,Abramchuk07}$^]$ The reason of this discrepancy lies presumably in an idealized perfectly regular form of the chain structures considered in our lattice approach. However, in reality the particles in MSEs are organised in ''wave-like'' irregular chains (see \ref{F15}), as was shown in the references.$^[$\cite{Jolly96,Coquelle05,Coquelle06}$^]$ The tensile deformation of a "wave-like" structure leads to effective shear deformation of the irregular chains, and the shear deformation as we have shown in this study results in the increase of the elastic modulus. Thus, irregularities in the chain-like distribution can possibly explain increase of the Young's modulus under application of the magnetic field.

In the present work we have considered the mechanical moduli of MSEs only for two deformational geometries: the shear deformation in the direction perpendicular to the magnetic field and the tensile deformation parallel to the magnetic field. We note here that in the case of chain-like and plane-like distributions of magnetic particles one deals with an anisotropic medium, which is characterized by a set of the mechanical moduli. For example, in the case of uniaxial media for a full characterization of the system one needs four independent moduli, that correspond to different geometries of the application of a small deformation with respect to the axis of anisotropy. Moreover, we note that the classical relationship between the tensile modulus ($E$), the shear modulus ($G$) and the Poisson's coefficient ($\nu$),
\begin{equation}
\nu =\frac{E}{2G}-1
\end{equation}
fails for uniaxial media, since both $E$ and $G$ depends on the geometry of a small deformation. Therefore, the uniaxial medium is characterized not by one but by three Poisson's coefficients,$^[$\cite{Schurmann05}$^]$ one of them can be even negative as was shown in the references.$^[$\cite{Dudek07,Dudek08}$^]$  Consideration of the mechanical moduli for other geometries and calculation of Poisson's coefficients can be a topic of further generalizations of our lattice approach.

\section{Conclusions} \label{S5}

In the present study we have developed a theory of mechanical behaviour of magneto-sensitive elastomers. We use a model in which magnetic particles are located in the sites of a regular rectangular lattice. Different distributions of particles in the space are considered: isotropic (the cubic lattice), chain-like and plane-like distributions. We show that interaction between the magnetic particles results in the contraction of an elastomer in the direction of the homogeneous magnetic field ($\varepsilon_{\rm eq}<0$) for all structures considered. Similar to the previous studies,$^[$\cite{Jolly96,Jolly96a,Davis99,Zhu06}$^]$ we show that the shear modulus $G$ increases for all types of distribution of magnetic particles with increasing magnetic field. On the other side, we show for the first time that in the frame of lattice approach the Young's  modulus $E$ decreases for the chain-like distribution and increases for the plane-like distribution of magnetic particles with increasing magnetic field. The shear modulus $G$ and the Young's  modulus $E$ are calculated at the minimum of free energy, where $\varepsilon=\varepsilon_{\rm eq}$. Thus, we take into account the magnetostriction effect, which is neglected upon calculation of the modulus in the previous studies.$^[$\cite{Jolly96,Jolly96a,Davis99,Zhu06}$^]$

\acknowledgement

This work was supported by funds of European Union and the Free State of Saxony.

\bibliography{bibliography}

\newpage

\end{document}